\documentclass[aps,twocolumn,pra,superscriptaddress,amsmath,showpacs,tightenlines,pdflatex]{revtex4}

\usepackage{amsmath}

\usepackage{graphicx,times}
\usepackage{amstext}
\usepackage{amsmath}            %serve per le subequazioni
\usepackage{amssymb}            %serve per il simbolo "marchio registrato", \circledR
\usepackage{graphicx}           %serve per le figure eps, ps etcyy
\usepackage{latexsym}
\usepackage{bm}

 \def\extra#1{\emph{#1}}

\def \etal{\textit{et al.}}
\newcommand{\tr}{{\rm tr\,}}

\newcommand{\job}{J. Opt. B~}

\begin{document}

\title{Two-photon and three-photon blockades in driven nonlinear systems}

\author{Adam Miranowicz}
\affiliation{Faculty of Physics, Adam Mickiewicz University,
PL-61-614 Pozna\'n, Poland} \affiliation{Advanced Science
Institute, RIKEN, Wako-shi, Saitama 351-0198, Japan}

\author{Ma\l{}gorzata Paprzycka}
\affiliation{Faculty of Physics, Adam Mickiewicz University,
PL-61-614 Pozna\'n, Poland}

\author{Yu-xi Liu}
\affiliation{Institute of Microelectronics, Tsinghua University,
Beijing 100084, China} \affiliation{Tsinghua National Laboratory
for Information Science and Technology (TNList), Tsinghua
University, Beijing 100084, China} \affiliation{Advanced Science
Institute, RIKEN, Wako-shi, Saitama 351-0198, Japan}

\author{Ji\v r\'\i\ Bajer}
\affiliation{Department of Optics, Palack\'{y} University, 772~00
Olomouc, Czech Republic}

\author{Franco Nori}
\affiliation{Advanced Science Institute, RIKEN, Wako-shi, Saitama
351-0198, Japan} \affiliation{Physics Department, The University
of Michigan, Ann Arbor, Michigan 48109-1040, USA}

\date{\today}

%------------------------------------------------------------------
\begin{abstract}
Photon blockade, in analogy to Coulomb's or phonon blockades,
is a phenomenon when a single photon in a nonlinear cavity
blocks the transmission of a second photon. This effect can
occur in Kerr-type systems driven by a laser due to strong
nonlinear photon-photon interactions. We predict the
occurrence of higher-order photon blockades where the
transmission of more than two photons is effectively blocked
by single- and two-photon states. This photon blockade can be
achieved by tuning the frequency of the laser driving field
to be equal to the sum of the Kerr nonlinearity  and the
cavity resonance frequency. We refer to this phenomenon as
two-photon blockade or two-photon state truncation via
nonlinear scissors, and can also be interpreted as
photon-induced tunneling. We also show that, for a
driving-field frequency fulfilling another resonance
condition and for higher strengths of the driving field, even
a three-photon blockade can occur but less clearly than in
the case of single- and two-photon blockades. We demonstrate
how various photon blockades can be identified by analyzing
photon-number correlations, coherence and entropic
properties, Wigner functions, and spectra of squeezing. We
show that two- and three-photon blockades can, in principle,
be observed in various cavity and circuit quantum
electrodynamical systems for which the standard single-photon
blockade was observed without the need of using higher-order
driving interactions or Kerr media exhibiting higher-order
nonlinear susceptibility.
\end{abstract}


\maketitle \pagenumbering{arabic}

%------------------------------------------------------------------
\section{Introduction}

In nonlinear optical systems driven by a coherent classical light
in a cavity, a single photon can impede the transmission of other
photons. This phenomenon is referred to as \emph{photon blockade}
(PB)~\cite{Imamoglu97}, in close analogy to the phenomenon of
\emph{Coulomb's blockade}~\cite{Kastner92}, where electron
transport is blocked by a strong Coulomb interaction in a confined
structure. In PB the next photon can enter the cavity only if the
first photon has left it; thus, a sequence of single photons can
be generated and such system can act as a single-photon turnstile
device. Evidently, the PB changes classical light into highly
nonclassical light exhibiting, in particular, photon antibunching
and sub-Poisson photon-number statistics.

The PB can be interpreted as nonlinear \emph{optical-state
truncation} or \emph{nonlinear quantum
scissors}~\cite{Leonski94,Miran01}, since the
infinitely-dimensional Fock-state expansion of a classical
driving field is truncated at the single-photon Fock's state.
The required nonlinearity can be induced by a strong
interaction between the cavity and the two-level (natural or
artificial) atom.

The PB has been predicted in various setups in cavity quantum
electrodynamics (QED)~\cite{Tian92,Werner99,Brecha99,Rebic99,
Rebic02,Kim99,Smolyaninov02}, and recently also in circuit
QED~\cite{Hoffman11,Lang11}. The PB was first demonstrated
experimentally with a single atom trapped in an optical
cavity~\cite{Birnbaum05}. This experiment was considered ``a
landmark event in the field of quantum optics and laser
science''~\cite{comment}. In solid-state systems, the PB was
experimentally demonstrated with a quantum dot in a photonic
crystal cavity~\cite{Faraon08} and with a single superconducting
artificial atom coupled to a microwave transmission-line resonator
(superconducting ``cavity'')~\cite{Hoffman11,Lang11}. The PB was
also predicted in quantum optomechanical systems~\cite{Rabl11}. An
analogous phenomenon of \emph{phonon blockade} was predicted for
an artificial superconducting atom coupled to a nanomechanical
resonator~\cite{Liu10}. The phonon blockade can be detected via
the PB if this system is additionally coupled with, e.g., a
superconducting microwave cavity~\cite{Didier11}.

The PB can have applications in quantum state engineering for the
controllable generation of a train of single photons exhibiting
highly nonclassical photon statistics. This suggests that the PB
(together with phonon blockades) can also be used as an indicator
of nonclassicality of mechanical systems~\cite{Liu10,Didier11}. As
another example of quantum engineering, the PB was also studied in
the context of cavity electromagnetically induced transparency in,
e.g., Refs.~\cite{Werner99,Rebic02}. The possibility of tunable
(by a classical driving field) transmission from photon blockade
to photon transparency in circuit-QED systems was described in
Ref.~\cite{Liu12}.

Moreover, single-photon tunneling, in close analogy to
single-electron tunneling, was experimentally observed during
light transmission through individual subwavelength
pinholes~\cite{Smolyaninov02}. This effect was explained in
terms of the PB, analogously to the Coulomb blockade
demonstrated in single-electron tunneling. A close analogy
between the PB and Coulomb blockade was also analyzed in
Ref.~\cite{Liu12} by presenting, e.g., a photonic analog of
the Coulomb staircase. Such studies shed more light on
quantum simulations of condensed-matter phenomena via optical
effects and vice versa.

In this paper, we show the possibility of observing the
multiphoton blockade in the standard cavity QED and circuit
QED systems, where the single-photon blockade was already
observed. These systems consist of a Kerr nonlinearity in a
cavity driven by a classical weak field. We will show how to
change the intensity and frequency of the driving field for a
given cavity frequency and the Kerr nonlinearity in order to
observe the two- and three-PBs.

Those multi-PBs can also be explained as \emph{photon-induced
tunneling.} That is, when there is a photon inside the
cavity, then the second or third photon can be absorbed by
the cavity via two-photon or three-photon processes.

Effects leading to multi-PB were already studied in the
literature as generalizations of the single-PB. For example,
several systems based on higher-order parametric driving
processes and/or higher-order Kerr nonlinearity were analyzed
in Refs.~\cite{Leonski96,Miran96,Leonski97} for the
observation of the nonstationary-field multi-PB. In the
Conclusions, we will briefly discuss the formal differences
and crucial experimental advantages of our approach over
these methods.

There is another approach to the PB, which is based on linear
systems, while the nonlinearity is induced by measurements.
This method is usually referred to as the linear optical
state truncation or linear scissors~\cite{Pegg98}. The
multiphoton state truncations (or multi-PBs) implemented via
linear scissors were studied in
Refs.~\cite{Koniorczyk00,Miran05}, as a generalization of the
single-photon state truncation~\cite{Pegg98,Ozdemir01}. It is
worth stressing that the PBs studied in this paper are based
on completely different principles and resources. We use, as
in the original single-PB
proposals~\cite{Imamoglu97,Leonski94}, nonlinear systems
(which cause nonlinear photon-photon interactions) without
measurements. Thus, the method used here can be referred to
as nonlinear scissors~\cite{Miran01}.

The paper is organized as follows. In Sec.~II, we explain the
occurrence of the single-PB and describe the method to
observe blockades up to two and three photons. In Sec.~III,
we demonstrate analytically the occurrence of the two-PB in
comparison to the single-PB. In Sec.~IV, we discuss various
signatures of the two-PB revealed by the photon-number
statistics, entropies, Wigner functions and spectra of
squeezing. We summarize our main results in the concluding
section.

%------------------------------------------------------------------
\begin{figure}

\includegraphics[width=.48\textwidth]{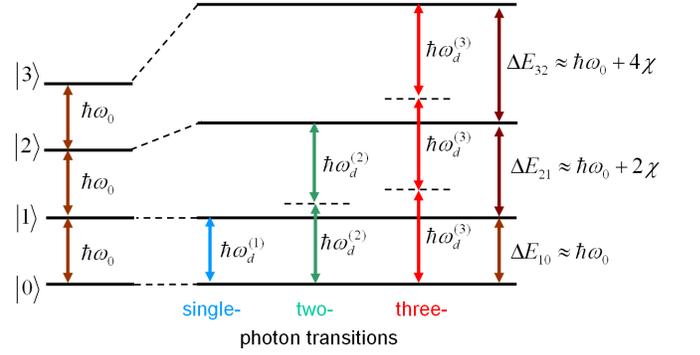}

\vspace*{-4mm} \caption{(Color online) Schematic energy-level
diagram explaining the occurrence of the $k$-photon blockade (and
the $k$-photon-induced tunneling) in terms of the $k$-photon
transitions induced by the driving field satisfying the resonance
condition $\Delta_k=0$, which corresponds to the driving-field
frequency $\omega_{d}=\omega^{(k)}_{d}=\omega_{0}+\chi(k-1)$. Due
to the Kerr-type nonlinearity (induced by, e.g., a qubit), the
cavity-mode levels $E_n^{(0)}=n\hbar\omega_{0}$ (shown on the
left) become non-equidistant as $\Delta
E_{n+1,n}=E_{n+1}-E_{n}\neq\;$const, where $E_n\approx
n[\hbar\omega_{0}+(n-1)\chi]$ (for $n=0,1,...)$ are the
eigenvalues of the Hamiltonian $\hat H$, given by Eq.~(\ref{H0}),
assuming $\epsilon\ll \chi$.}
\end{figure}
%------------------------------------------------------------------
\begin{figure}

\includegraphics[width=.46\textwidth]{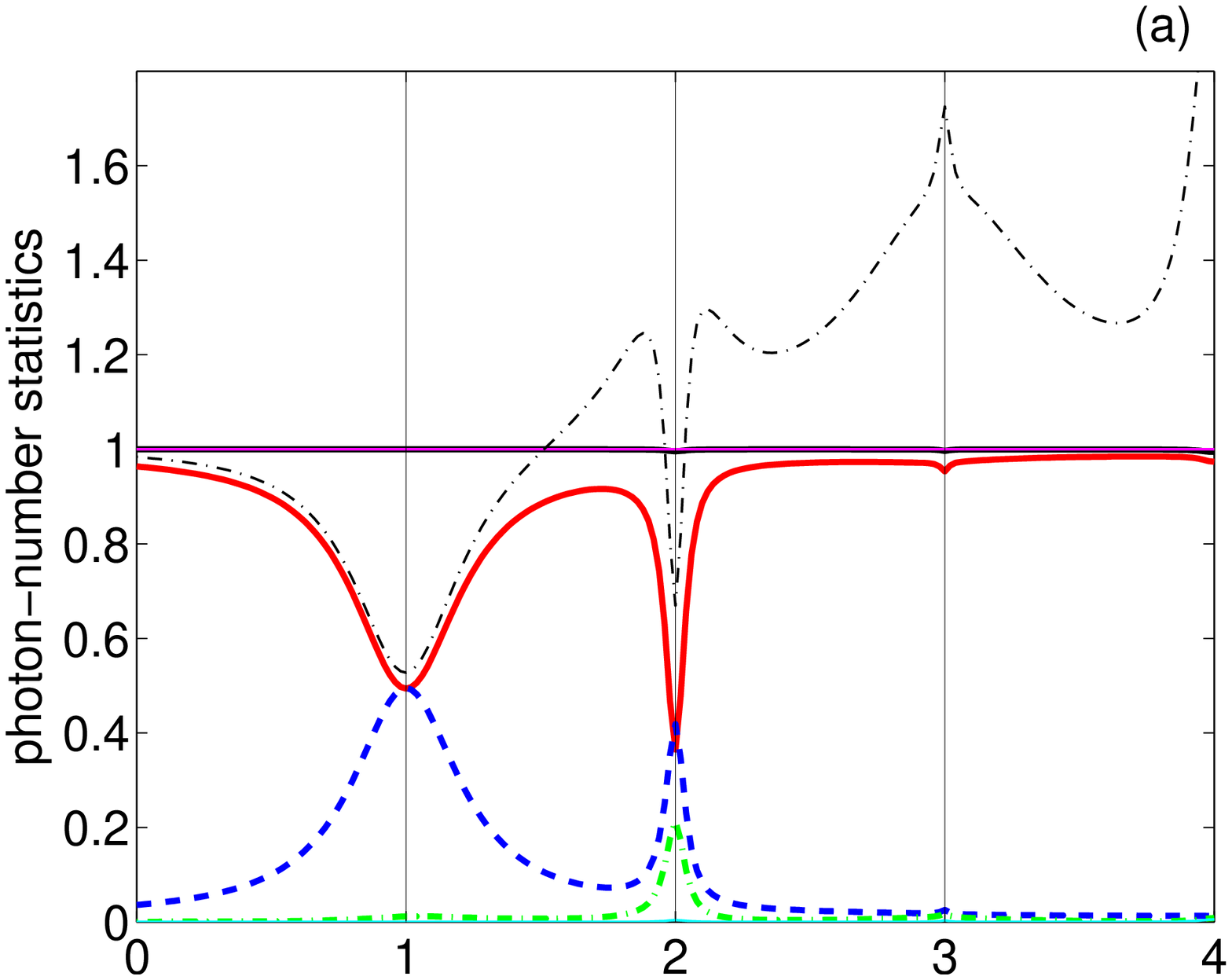}

%\vspace*{-5mm}

\includegraphics[width=.46\textwidth]{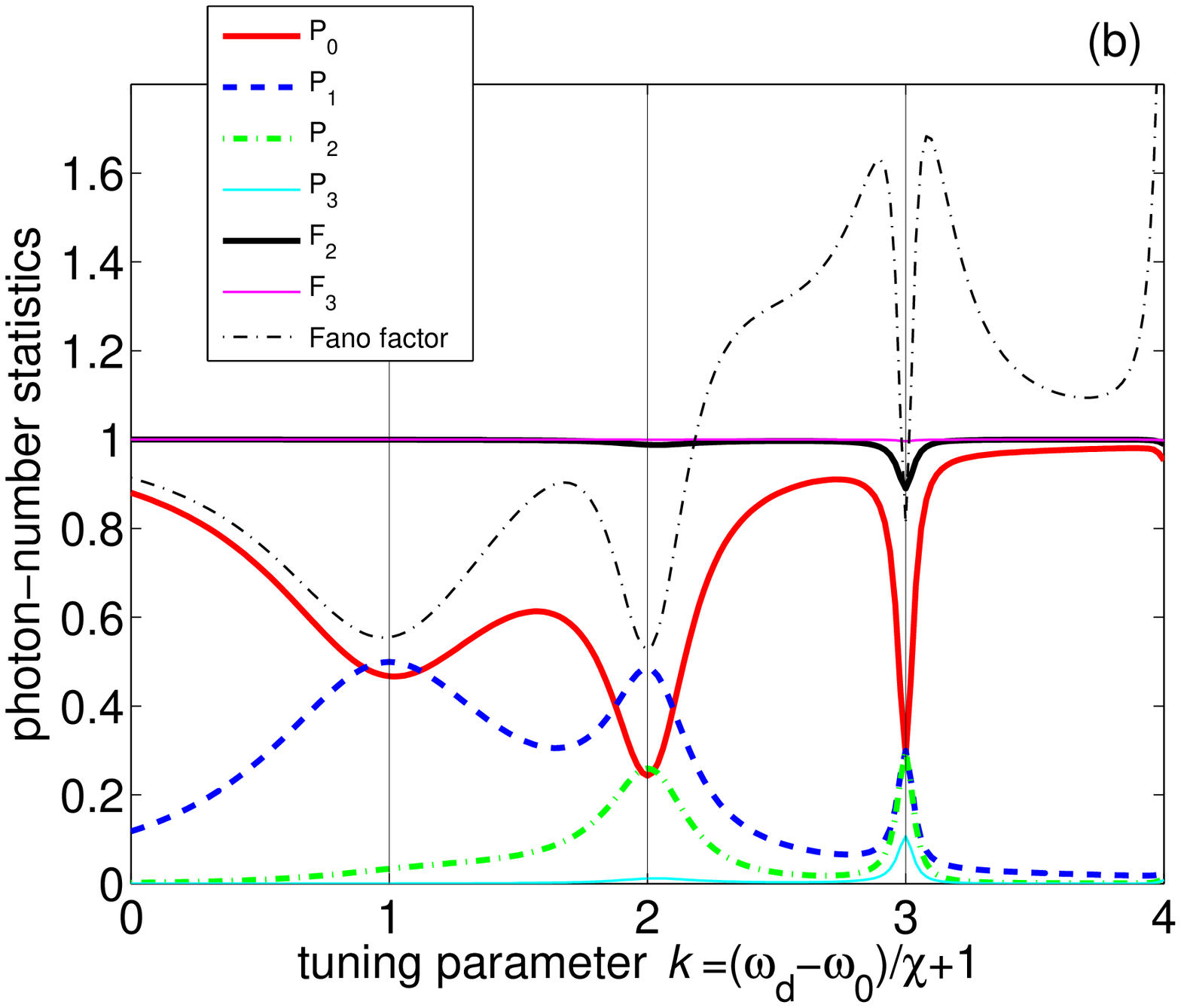}

\vspace*{-4mm} \caption{(Color online). Photon-number
probabilities  $P_n=\langle n|\hat\rho_{\rm ss}|n\rangle$ for
the steady-state solutions $\hat\rho_{\rm ss}$ of the master
equation for the Hamiltonian
$\hat{H}_{\mathrm{rot}}^{(k)}(0)$ as a function of the tuning
parameter $k$ assuming the driving strengths (a)
$\epsilon=5\gamma$ and (b) $\epsilon=11.56\gamma$. Moreover,
we assume: the Kerr nonlinearity $\chi=30\gamma$, the damping
constant $\gamma=1$, and the mean number of thermal photons
$\bar{n}_{\rm th}=0.01$. Figures also show the truncation
fidelity $F_m$ (for $m=2,3$) and the Fano factor ${F}$. Note
that the field exhibits sub-Poisson (super-Poisson)
photon-number statistics if ${F}<1$ (${F}>1$). It is seen in
both figures that resonances at $k=1$ and 2 can be
interpreted as the single- and two-photon blockades,
respectively. But the three-photon blockade at $k=3$ is
apparent in (b) only.}
\end{figure}
%------------------------------------------------------------------
\begin{figure}
\includegraphics[width=.24\textwidth]{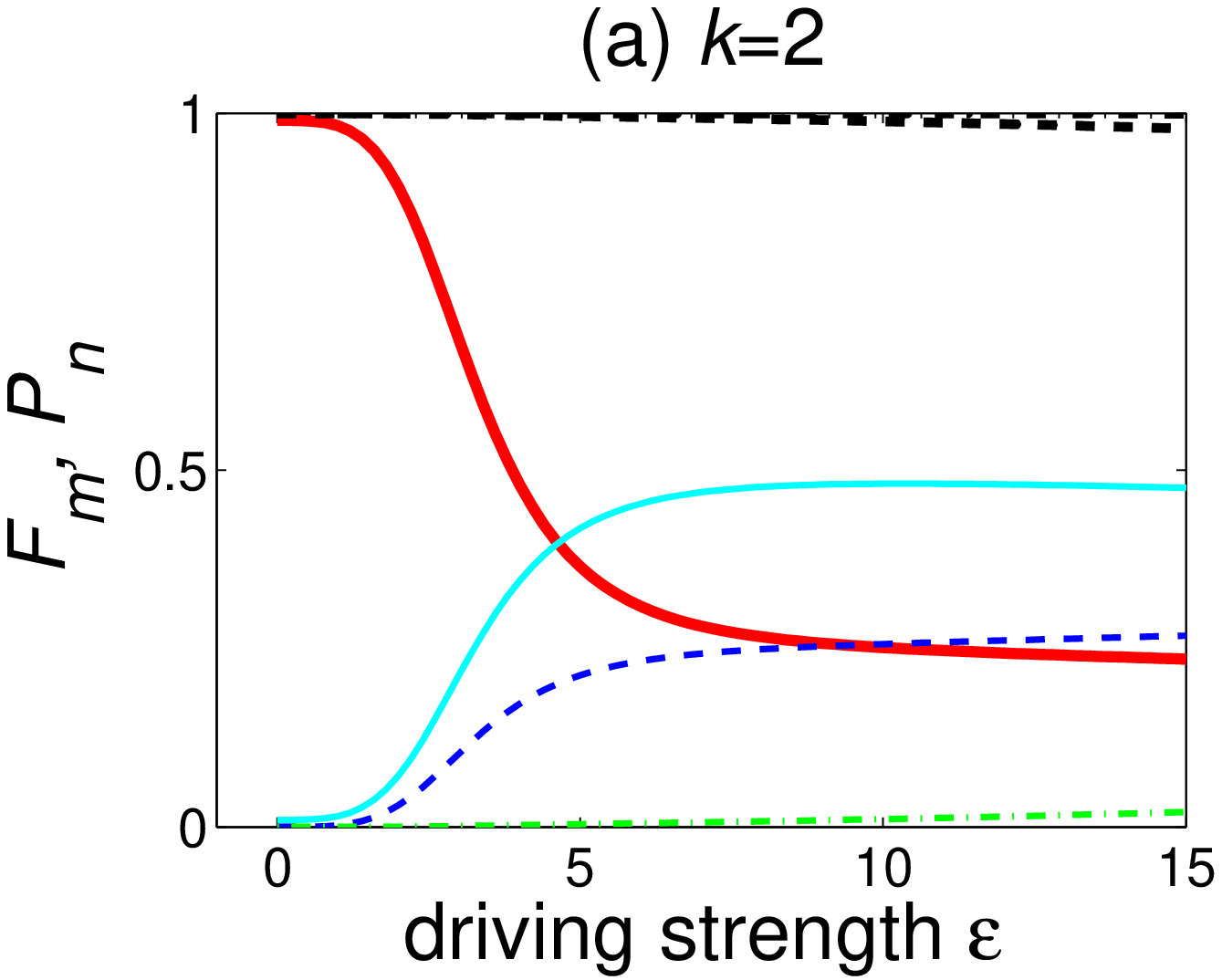}
\hspace*{-2mm}
\includegraphics[width=.24\textwidth]{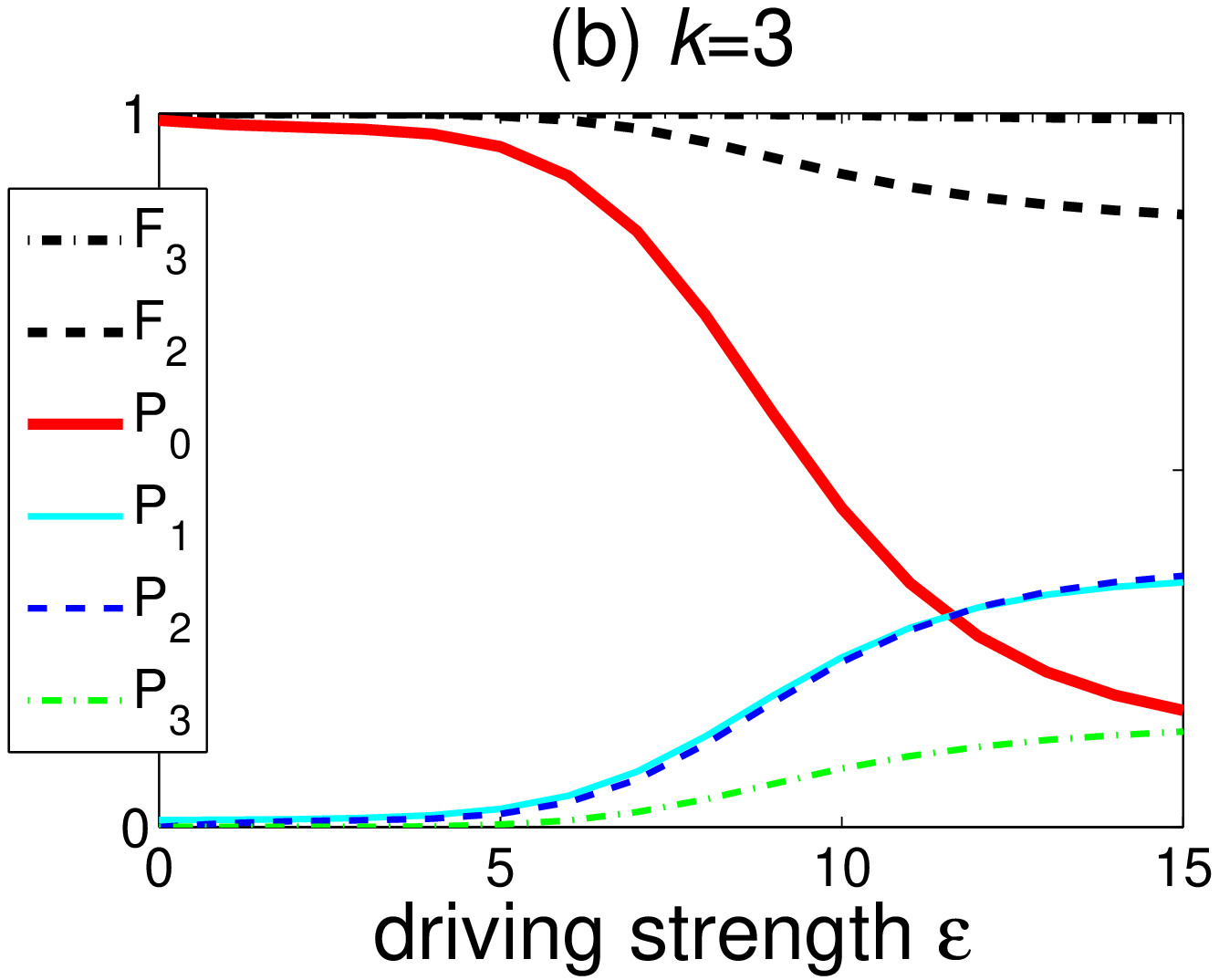}

\vspace*{-4mm} \caption{(Color online). Photon-number
probabilities $P_n=\langle n|\hat\rho_{\rm ss}|n\rangle$ and
fidelities $F_m=\sum_{n=0}^m P_n$ of the $m$-photon
truncation for the Hamiltonian
$\hat{H}_{\mathrm{rot}}^{(k)}(0)$ for the resonances at (a)
$k=2$ and (b) $k=3$ as a function of the driving strength
$\epsilon$ assuming the Kerr nonlinearity $\chi=30\gamma$,
the damping constant $\gamma=1,$ and the mean thermal-photon
number $\bar{n}_{\rm th}=0.01$. Note that $P_0\approx
P_1\approx P_2$ for the driving strength $\epsilon=11.56$ and
$k=3$. This value of $\epsilon$ is chosen in Figs.~2(b), 5, 8
(for $k=3$), and 11. Also note that in (a) for $k=2$, there
are two crossings, one for $P_0=P_1$ and another one for
$P_0=P_2$.}
\end{figure}

%------------------------------------------------------------------
\section{From single-photon to multiphoton blockades}

\subsection{Hamiltonians}

Photon-blockade effects can be observed in a cavity with a
nonlinear medium coherently driven by a laser field as described
by the following effective
Hamiltonian~\cite{Imamoglu97,Leonski94}:
\begin{equation}
\hat{H}=\hbar\omega_{0}\hat{a}^{\dagger}\hat{a}
+\hbar\chi(\hat{a}^{\dagger})^{2}\hat{a}^2+\hbar\epsilon(\hat{a}
e^{i\omega_{d}t}+\hat{a}^{\dagger}e^{-i\omega_{d}t}), \label{H0}
\end{equation}
where  $\hat{a}$ $(\hat{a}^{\dagger})$ denotes the annihilation
(creation) operator, $\omega_{0}$ is the resonance frequency of
the cavity, $\omega_{d}$ is the driving laser frequency, $\chi>0$
is the Kerr nonlinearity, i.e., the photon-photon interaction
strength proportional to the real part of the third-order
nonlinear susceptibility Re$(\chi^{(3)})$, and $\epsilon>0$ is the
driving strength (the Rabi frequency of the laser).

Equation~(\ref{H0}) presents an effective Hamiltonian, which can
be obtained from various microscopic Hamiltonians describing a
variety of systems. For example, one can analyze a quantum
two-level system (qubit) off-resonantly coupled to a driven
cavity. This system can be described in the rotating-wave
approximation by the following Hamiltonian:
\begin{eqnarray}
 \hat H &=& \frac12 \hbar\omega\hat\sigma_z+\hbar\omega'_{0}\hat{a}^{\dagger}\hat{a}
+\hbar g(\hat\sigma^{+}\hat a + \hat a^{\dag}\hat\sigma^{-})
\\\nonumber &&+\hbar\epsilon(\hat{a}
e^{i\omega_{d}t}+\hat{a}^{\dagger}e^{-i\omega_{d}t}), \label{JCM}
\end{eqnarray}
where the first three terms correspond to the standard
Jaynes-Cummings Hamiltonian and the last term, as in
Eq.~(\ref{H0}), describes the interaction between the quantum
cavity mode and the classical driving field with strength
$\epsilon$. Moreover, $\sigma_{z}=|e\rangle\langle
e|-|g\rangle\langle g|$ denotes the Pauli operator,
$\sigma^{+}=|e\rangle\langle g|$ ($\sigma^{-}=|g\rangle\langle
e|$) is the qubit raising (lowering) operator, $|g\rangle$
($|e\rangle$) is the ground (excited) state of the qubit, $\omega$
is the qubit transition frequency, and $\omega'_{0}$ is the
resonance frequency of the cavity in Eq.~(\ref{JCM}). Moreover,
other quantities are explained below Eq.~(\ref{H0}). This
Hamiltonian in the dispersive approximation, where the qubit
remains in its ground state, can be reduced to the Hamiltonian,
given by Eq.~(\ref{H0}) (for a derivation see, e.g.,
Refs.~\cite{Boissonneault09,Liu10} in the circuit QED context).

The unitary operation
$\hat{U}=\exp(-i\omega_{d}\hat{a}^{\dagger}\hat{a}t)$ transforms
the Hamiltonian~(\ref{H0}) into
\begin{equation}
\hat{H}_{\mathrm{rot}}=\hat{U}^{\dagger}\hat{H}\hat{U}-i\hbar\hat{U}^{\dagger}\frac{\partial}{\partial
t}\hat{U}.\label{Hrot}
\end{equation}
Thus, in the rotating frame, one obtains the following
time-independent Hamiltonian,
\begin{equation}
\hat{H}_{\mathrm{rot}}^{(1)}
=\hbar\Delta_{1}\hat{a}^{\dagger}\hat{a}
+\hbar\chi(\hat{a}^{\dagger})^{2}\hat{a}^2+\hbar\epsilon(\hat{a}+\hat{a}^{\dagger}),\label{H1}
\end{equation}
where $\Delta_{1}=\omega_{0}-\omega_{d}$. The single-PB can be
observed in the resonant case $\omega_{d}=\omega_{0}$ assuming
that the driving strength $\epsilon$ is much smaller than the Kerr
nonlinearity $\chi$. This effect can be interpreted as the
single-photon Fock state blockading the generation of two or more
photons.

We observe that Hamiltonian~(\ref{H1}) can be rewritten as
follows:
\begin{equation}
\hat{H}_{\mathrm{rot}}^{(k)}(\Delta_{k})
=\hbar\Delta_{k}\hat{a}^{\dagger}\hat{a}
+\hbar\chi\hat{a}^{\dagger}\hat{a}(\hat{a}^{\dagger}\hat{a}-k)
+\hbar\epsilon(\hat{a}+\hat{a}^{\dagger}),\label{Hk}
\end{equation}
where the frequency mismatch is
\begin{equation}
\Delta_{k}=\omega_{0}+\chi(k-1)-\omega_{d}\label{Delta_k}
\end{equation}
for some positive $k$. For convenience, we shall refer to $k$ as a
tuning parameter. In the special case of $k=1,$ this Hamiltonian
reduces to the standard form, given by Eq.~(\ref{H1}).

%------------------------------------------------------------------
\subsection{Steady states as a function of the tuning parameter $k$}

The evolution of the system, given by Eq.~(\ref{Hk}), for the
reduced density operator $\hat \rho(t)$ under Markov's
approximation, can be governed by the standard master
equation~\cite{Carmichael}, which for the steady state
$\hat{\rho}_{{\rm ss}}=\hat \rho(t\rightarrow \infty)$ is given by
\begin{eqnarray}
&0= -\frac{i}{\hbar}[ \hat{H}_{\rm rot}^{(k)}
 (\Delta_{k}),\hat{\rho}_{{\rm ss}}]
 +\frac{\gamma}{2}\bar{n}_{\rm th}(2\hat a^{\dagger}\hat{\rho}_{{\rm ss}} \hat a -\hat a\hat
 a^{\dagger}\hat{\rho}_{{\rm ss}}\hspace{9mm}
 \nonumber \\ &
 -\hat{\rho}_{{\rm ss}} \hat a\hat a^{\dagger})+\frac{\gamma}{2}(\bar{n}_{\rm th}+1)(2\hat a\hat{\rho}_{{\rm ss}} \hat a^{\dagger}
 -\hat a^{\dagger}\hat a\hat{\rho}_{{\rm ss}}-\hat{\rho}_{{\rm ss}} \hat a^{\dagger}\hat a),
 \label{ME}
\end{eqnarray}
where $\gamma$ denotes the damping constant, $\bar{n}_{\rm
th}=\{\exp[\hbar\omega/(k_{B}T)]-1\}^{-1}$ is the mean number of
thermal photons, $k_{B}$ is the Boltzmann constant, and $T$ is the
reservoir temperature at thermal equilibrium. The master equation
can be given in terms of a Liouvillian superoperator and the
steady-state solution $\hat \rho(t)$ can be obtained by applying,
e.g., the inverse power method~\cite{Tan}.

The single-PB can occur if the conditions
\begin{eqnarray}
\gamma\ll\epsilon\ll\chi
 \label{conditions}
\end{eqnarray}
are satisfied. Hereafter, we only analyze the resonant case
$\Delta_{k}=0$, which is related to the resonant $k$-photon
transitions shown in Fig.~1. This condition implies that the
tuning parameter $k$ is related to the Kerr nonlinearity, and
the driving-field, and cavity frequencies as follows:
\begin{eqnarray}
k=(\omega_{d}-\omega_0)/\chi+1.
 \label{k}
\end{eqnarray}

Figure~2 shows how the photon-number probabilities $P_n$ of the
steady states depend on the tuning parameter $k$. By analyzing
this figure, one can discover  various kinds of PB effects, which
appear not only for the standard resonant case of $k=1$ but also
for $k\neq1$. For example, in the special case of $k=2$, a
higher-order effect occurs with at most two photons effectively
generated in the system. We refer to this effect as the
\emph{two-photon blockade}, which means that the single- and
two-photon Fock states blockade the generation of more photons.

Figure~2(b), which is similar to Fig.~2(a) but obtained
assuming larger driving strength $\epsilon$, clearly shows a
\emph{three-photon blockade} for $k=3$, which refers to a
phenomenon where the Fock states $|m\rangle$ (for $m=1,2,3$)
blockade the transmission of the Fock states $|n\rangle$ with
higher photon number (i.e., $n>3$). Note that, in Fig.~2(a)
for $k=3$, the probabilities of generating
$|1\rangle,\;|2\rangle$ and $|3\rangle$ are nonzero but are
much lower than the probability of observing the vacuum
state. So, this behavior for the parameters of Fig.~2(a),
contrary to Fig.~2(b), cannot be considered a genuine
three-PB.

It is also quite evident why it is necessary to increase the
strength of the driving field to obtain a good-quality three-PB
since we need a larger average photon number. For weak driving,
the average photon number of the driving field is less than three
photons, thus this is not enough to induce a three-photon
transition as shown in Fig. 1.

In Fig.~2, in particular, we show the fidelity of the $m$-photon
truncation defined as
\begin{equation}
F_m(\hat{\rho}_{{\rm ss}})=\sum_{n=0}^m P_n=\sum_{n=0}^m\langle
n|\hat{\rho}_{{\rm ss}}|n\rangle.\label{fidelity}
\end{equation}
We refer to the $m$-PB if the truncation fidelity $F_m\approx 1$
and $F_n\ll 1$ for $n<m$. In Fig.~2(a), it is seen that
$F_1\approx 1$ at $k=1$ ($F_2\approx 1$ at $k=2$) corresponding to
the single-PB (two-PB). But the resonance at $k=3$, for the chosen
driving strength, can hardly be considered the three-PB for the
driving strength $\epsilon=5\gamma$. By contrast, $F_3\approx 1$
at $k=3$ for the much higher driving strength
($\epsilon=11.56\gamma$), as shown in Fig.~2(b). Thus, we
interpret the latter case as the true three-PB, as already
mentioned.

By analyzing the dependence of the photon-number probabilities
$P_n=\langle n|\hat\rho_{\rm ss}|n\rangle$ on the tuning parameter
$k$, as shown in Fig.~2(a), we observe that there is a clear dip
in the vacuum-state probability $P_0$ and a peak in the
single-photon probability $P_1$ at $k=1$ and 2. The two-photon
probability $P_2$ is nonzero only near $k=2$. The peak maximum in
$P_2$ does not occur at $k=1$ but it is evident at $k=2$, which
implies that the single-photon truncation fidelity $F_1$ differs
from 1 near $k=2$. By contrast, the two-photon truncation fidelity
$F_2$ on the scale of Fig.~2(a) is practically equal to one. Thus,
the contribution of Fock's states with three or more photons is
negligible.  For these reasons, we refer to the two-PB at $k=2$.
One can see in Fig.~2(a) a very slight dip in $P_0$ at $k=3$,
which can be, however, much deeper for the approximately
twice-larger driving strength $\epsilon$, as shown in Fig.~2(b).
The dip in $P_0$ at $k=3$ is accompanied by the clear appearance
of the peaks of $P_n$ for the photon numbers $n=1,2,3$. It is seen
that for $k=3$ (but also close to this point), the three-photon
truncation fidelity $F_3\approx 1$ contrary to $F_2\ll 1$. This
explains why we refer to this effect as the three-PB at $k=3$ for
a suitably large driving strength $\epsilon$ as, e.g., in
Fig.~2(b).

Figure 3 shows how by increasing the driving strength
$\epsilon$ one can change the steady-state probabilities of
the generation of $n$-photons for chosen values of the Kerr
nonlinearity, damping constant, and for the resonance
conditions (a) $k=2$ and (b) $k=3$. We also depicted the two-
and three-photon truncation fidelities, which practically
equal to 1 for all the values of $\epsilon$ in Figs. 3(a) and
3(b), respectively. It is seen that for the small driving
strength one cannot generate photons in the cavity in the
steady-state limit, as $P_0\approx 1$. This is a trivial
case, which can be interpreted as the zero-PB. For larger
values of $\epsilon$, the true single- and two-PB effects are
observed. In particular, it is seen in Fig.~3(b) that for the
driving strength $\epsilon=11.56$ the probabilities $P_n$ of
observing $n=0,1,2$ photons are approximately the same. The
three-photon probability is smaller but still nonzero.  We
have chosen this $\epsilon$ to analyze various properties of
the three-PB in Fig.~2 and other figures.

%------------------------------------------------------------------
\section{Analytical description of photon blockades}

%------------------------------------------------------------------
\subsection{Steady-state photon blockade}

Here, we explain analytically the standard (i.e., steady-state) PB
effects, i.e., by including dissipation in the system in the
infinite-time limit.

We start from the infinite-dimensional Hamiltonian, given by
Eq.~(\ref{Hk}) in the resonant case $\Delta_{k}=0$ for a given
$k$, and formally truncate it to a finite-dimensional Hilbert
space. For example, to show explicitly the two-PB ($k=2$), one
should analyze the Hamiltonian at least in the four-dimensional
Hilbert space:
\begin{equation}
\hat H_{\mathrm{trunc}}^{(2)}(0)=\left(\begin{array}{cccc}
0 & \epsilon & 0 & 0\\
\epsilon & -\chi & \sqrt{2}\epsilon & 0\\
0 & \sqrt{2}\epsilon & 0 & \sqrt{3}\epsilon \\
0 & 0 & \sqrt{3}\epsilon & 3\chi
\end{array}\right),\label{H23}
\end{equation}
which is given in the standard Fock basis. We have to show that
the contribution of three-photon terms $\langle3|\hat{\rho}_{{\rm
ss}}^{(2)}|n\rangle$ (for $n=0,1,2,3$) in the steady-state density
matrix is negligible.

The conditions given by Eq.~(\ref{conditions}) should be fulfilled
to observe a PB effect, so let us denote
\begin{eqnarray}
\frac{\gamma}{\epsilon}=\delta,\quad\frac{\epsilon}{\chi}=d\delta,
 \label{delta}
\end{eqnarray}
where $\delta\ll1$ and $d\delta\ll1$. To find an analytical
approximate steady-state solution for $\hat{\rho}_{{\rm
ss}}^{(2)}$, we substitute the truncated four-dimensional
Hamiltonian into the master equation, given by
Eq.~(\ref{ME}), and solve the set of equations for each of
the Fock-state elements of the density matrix $\langle
m|\hat{\rho}_{{\rm ss}}^{(2)}| n \rangle$, for $m,n=0,1,2,3$.
Then we expand these solutions in power series of $\delta$
and neglect terms proportional to $\delta^{2}$ or higher
powers. Thus, we find the following approximate steady-state
solution up to $\delta^1$:
\begin{equation}
\hat{\rho}_{{\rm
ss}}^{(2)}\approx\frac{1}{1+8d^{2}}\left(\begin{array}{cccc}
1+2d^{2} & x^{*}\delta & id\sqrt{2} & z^{*}\delta \\
x\delta & 4d^{2} &  y^*\delta  & 0\\
-id\sqrt{2} & y\delta & 2d^{2} & -\frac{2}{3}\sqrt{3}d^{3}\delta \\
z\delta & 0 & -\frac{2}{3}\sqrt{3}d^{3}\delta & 0\\
\end{array}\right),
\label{sol2a}
\end{equation}
where $x=-2d^{3}-2id^{2}+d$, $y=-\sqrt{2}d^{2}\left( 2d+i\right),$
and $z=\frac{1}{3}i\sqrt{6}d^{2}$.   Our numerical solutions
slightly differ from Eq.~(\ref{sol2a}) for the parameters chosen
in Fig.~2(a), but it is clearly seen that the contribution of the
three-photon terms $\langle 3|\hat{\rho}_{{\rm ss}}^{(2)}| 3
\rangle$ can be neglected,  contrary to the terms $\langle
m|\hat{\rho}_{{\rm ss}}^{(2)}| m \rangle$ with smaller number $m$
of photons. Also other elements $\langle 3|\hat{\rho}_{{\rm
ss}}^{(2)}| m \rangle$ are either equal to zero (for $m=1$) or
proportional to $\delta\ll 1$ (for $m=0,2$), so they can be
neglected. Thus, our solutions explain the two-PB.

For comparison, let us analyze the single-PB described by the
infinite-dimensional Hamiltonians,  given by Eqs.~(\ref{H1})
or~(\ref{Hk}) for $k=1$, but truncated to the three-dimensional
subspace, as given by
\begin{equation}
 \hat H_{\mathrm{trunc}}^{(1)}(0)=\left(\begin{array}{ccc}
0 & \epsilon & 0\\
\epsilon & 0 & \sqrt{2}\epsilon\\
0 & \sqrt{2}\epsilon & 2\chi
\end{array}\right).\label{H20}
\end{equation}
By performing calculations analogous to the former case, we
find the following steady-state solution up to $\delta^2$:
\begin{equation}
\hat{\rho}_{{\rm ss}}^{(1)}\approx\left(\begin{array}{ccc}
\frac{1}{2}+\frac{1}{16}(1-4d^{2})\delta ^{2} & x^{*}\delta & y^{*}\delta^2\\
x\delta & \frac{1}{2}-\frac{1}{16}\delta ^{2} & -\frac{1}{4}\sqrt{2}d\delta\\
y\delta^2 & -\frac{1}{4}\sqrt{2}d\delta & \frac{1}{4}d^{2}\delta
^{2}
\end{array}\right),
\end{equation}
where $x=-\frac{1}{4} \left( 2d+i\right)$ and
$y=\frac{1}{8}\sqrt{2}d\left( d+i\right)$, which  reduces to
\begin{equation}
\hat{\rho}_{{\rm
ss}}^{(1)}\approx\frac{1}{2}\left(\begin{array}{ccc}
1 & -(d -\frac{1}{2}i)\delta & 0\\
-(d +\frac{1}{2}i)\delta & 1 & -\frac{1}{2}d\sqrt{2}\delta\\
0 & -\frac{1}{2}d\sqrt{2}\delta & 0
\end{array}\right)
\label{sol1}
\end{equation}
assuming $\delta^2 \approx 0$. It is seen in Eq.~(\ref{sol1}) that
$\langle2|\hat{\rho}_{{\rm ss}}^{(1)}|n\rangle$ vanishes for
$n=0,2$ and is $\sim\delta\ll 1$ for $n=1$. So, the contribution
of all the two-photon states can be neglected, which clearly
explains the physical meaning of the single-PB for $k=1$. Note
that $\langle2|\hat{\rho}_{{\rm ss}}^{(2)}|2\rangle\neq 0$ as
described by Eq.~(\ref{sol2a}), which corresponds to the two-PB
for $k=2$.

%------------------------------------------------------------------
\begin{figure}
\includegraphics[width=.45\textwidth]{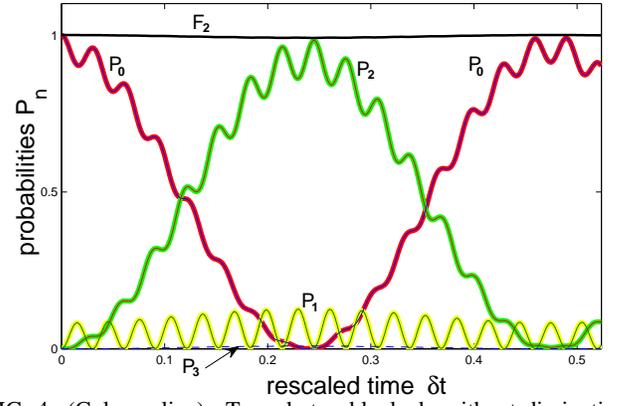}

\vspace*{-4mm} \caption{(Color online). Two-photon blockade
without dissipation: Evolution of the photon-number
probabilities $P_n$ for the tuning parameter $k=2$ according
to our precise numerical calculations (thin curves) in the 100-dimensional Hilbert space and
approximate solutions (thick curves),
given by Eqs.~(\ref{eig2})--(\ref{psi2}), obtained in the
truncated four-dimensional Hilbert space. Excellent agreement
of these solutions implies that the truncation fidelity
$F_2=P_0+P_1+P_2$ is almost exactly equal to one (as shown by
black line). This explains the meaning of the optical-state
truncation or the nonstationary-state two-photon blockade. It
is seen that the main contribution to $F_2$ is from the Fock
states $|0\rangle$ and $|2\rangle$, which corresponds to the
two-photon transitions shown in Fig. 1.}
\end{figure}
%------------------------------------------------------------------
\begin{figure}
\includegraphics[width=.45\textwidth]{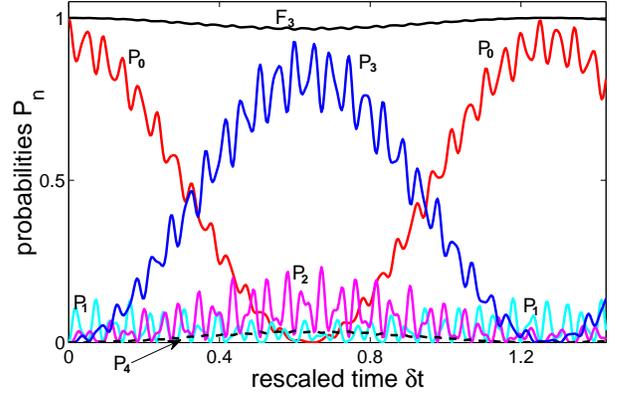}

\vspace*{-4mm} \caption{(Color online). Three-photon blockade
without dissipation: Evolution of the photon-number
probabilities $P_n$ as in Fig.~4 but for $k=3$ and all the
parameters (except $\gamma=0$) as in Fig.~2(b). It is seen
that the three-photon truncation fidelity $F_3$ slightly
deviates from one. For clarity, we present here only precise
solutions obtained in a large-dimensional Hilbert space. It
is evident that solely the Fock states $|0\rangle$ and
$|3\rangle$ are interchangeably highly populated, which
corresponds to the three-photon transitions shown in Fig.~1.
The probabilities $P_n$ for $n=1,2$ are relatively small but
not negligible, while $P_n$ for $n>3$ can practically be
ignored (e.g., $P_4$ is shown by broken curve). }
\end{figure}

%------------------------------------------------------------------
\subsection{Photon blockade without dissipation}

Here, we shortly describe nonstationary-state PB assuming no
dissipation. The main results of this section are summarized
in Figs. 4 and 5.

To describe the two-PB, let us formally confine the Hilbert
space of our system to four dimensions. Thus, we use the
Hamiltonian, given by Eq.~(\ref{H23}). Its exact eigenvalues
and eigenvectors can be calculated analytically, but they are
too lengthy to be presented here. Instead of this, we define
a small parameter $\delta\ll 1$ as the ratio of the driving
strength $\epsilon$ and the Kerr nonlinearity $\chi$ as given
in Eq.~(\ref{delta}) but, for simplicity, we assume here
$d=1$. Then, we find the power-series expansion in $\delta$
and keep the terms up to $\delta^2$ only. Thus, we find the
following eigenvalues:
\begin{eqnarray}
\lambda_{1}\approx\--\chi(3 \delta ^2+1),\quad
\lambda_{2}\approx\chi  \delta ^2x_{-},\hspace{5mm}
\nonumber \\
\lambda_{3}\approx\chi \delta^2x_{+},\quad
\lambda_{4}\approx\chi(\delta ^2+3),\hspace{8mm} \label{eig2}
\end{eqnarray}
and the corresponding eigenvectors:
\begin{eqnarray}
|\lambda_{1}\rangle&\approx&N_1[ 2 \sqrt{2} \delta |0\rangle - 2 \sqrt{2} \left(3 \delta ^2+1\right) |1\rangle \nonumber \\ && +4 \delta  |2\rangle - \sqrt{3} \delta ^2|3\rangle],\\
|\lambda_{2}\rangle&\approx&N_{_{-}}[(\delta ^2+3)  |0\rangle +3x_{_{-}} \delta   |1\rangle \nonumber \\ && +(x_{_{-}}  \delta ^2  -3)|2\rangle + \sqrt{3} \delta |3\rangle ],\\
|\lambda_{3}\rangle&\approx&N_{+}[(\delta ^2+3)  |0\rangle +3x_{+} \delta   |1\rangle \nonumber \\ && -(x_{_{+}}  \delta ^2  -3)|2\rangle - \sqrt{3} \delta |3\rangle ],\\
|\lambda_{4}\rangle&\approx&N_4[ \delta ^2|1\rangle+2 \sqrt{2}\delta
|2\rangle+2    \sqrt{6} |3\rangle ],
\label{eigv2}
\end{eqnarray}
where $x_{\pm} =1\pm\sqrt{2}$, and the normalization constants are
$N^{-2}_1\approx 8(1 + 9\delta^2)$, $N^{-2}_{\pm}\approx6(3+5\pm2
\sqrt{2} \delta ^2$), and $N^{-2}_4\approx8(3+\delta ^2)$. So,
assuming no damping and no photons initially in the cavity, the
time evolution of the truncated system can be given by
\begin{equation}
|\psi(t)\rangle=\sum_{j=1}^3\exp(-i\lambda_jt)\langle
\lambda_j|0\rangle |\lambda_j\rangle+{\cal O}(\delta^3).
\label{psi2}
\end{equation}
Note that, since the initial state is $|0\rangle$, there is no
contribution of $|\lambda_{4}\rangle $. After substituting
Eqs.~(\ref{eig2})--(\ref{eigv2}) into Eq.~(\ref{psi2}), we find
that
\begin{eqnarray}
\langle3|\psi(t)\rangle&=&\frac{\sqrt{6}}{2}\delta\left(e^{-i\lambda_2t}-e^{-i\lambda_3t}\right)+{\cal
O}(\delta^3)\nonumber\\
&=&\frac{\delta^3}{\sqrt{3}}(2-3e^{it}+i6t)+{\cal
O}(\delta^3)={\cal O}(\delta^3), \label{c3}
\end{eqnarray}
which shows explicitly the negligible contribution from the
three-photon Fock state in the generated state $|\psi(t)\rangle$.
Analogously we can show no important contribution from the Fock
states with more photons. Thus, we conclude the occurrence of
two-PB in the dissipation-free system.

In Fig.~4, we compare (i)
the analytical approximate solution, given by Eq.~(\ref{psi2}),
with  numerical solutions obtained in the Hilbert spaces of
dimension (ii) $N_{\rm dim}=100$ (which effectively corresponds to
$N_{\rm dim}=\infty$) and (iii) $N_{\rm dim}=4$  (without applying
expansions in the power series of $\delta$) for some parameters
satisfying the conditions (\ref{conditions}). On the scale of
Fig.~4, there is apparently no difference between the solutions
(ii) and (iii) at all, and very tiny discrepancy between them and
the analytical approximate solution (i) for the relatively large
$\delta=1/6$. This excellent agreement between the approximate and
precise solutions, convincingly demonstrate the blockade up to the
two-photon state $|2\rangle$. It is worth stressing that, although
the contribution of the three-photon state $|3\rangle$ is very
small, the calculations have to be performed in the Hilbert space
including the state $|3\rangle$.

For comparison, we recall that the truncated three-dimensional
Hamiltonian $\hat H_{\mathrm{trunc}}^{(1)}(0)$ under standard
conditions, given by Eq.~(\ref{conditions}), leads to the
following simple evolution~
\begin{eqnarray}
  |\psi(t)\rangle = \cos(\epsilon t)|0\rangle
  -i \sin(\epsilon t)|1\rangle+{\cal O}(\delta^2),
  \label{psi1}
\end{eqnarray}
assuming $|\psi(0)\rangle=|0\rangle$. As shown by Leo\'nski and
Tana\'s in Ref.~\cite{Leonski94}, this solution well approximates
the precise numerical evolution of the infinite-dimensional
system. This effect we refer to as the nonstationary-state
single-PB, but it is usually called  the single-photon optical
truncation~\cite{Miran01}.

Finally, we note that these PB effects can also be interpreted as
\emph{photon-induced tunneling} studied in the context of PB in,
e.g., Refs.~\cite{Smolyaninov02,Faraon08,Majumdar12}. An
especially simple interpretation can be found for the
dissipation-free PBs. Specifically, we can say that the PB effect
for $k=2$, as shown in Fig.~4, mainly corresponds to the
\emph{dominant} transition from the ground to the second-excited
state with the two-photon resonant transition condition as
schematically presented in Fig.~1. Analogously, the PB effect for
$k=3$, as shown in Fig.~5, describes mainly the three-photon
resonant transition from the ground to the third-excited state,
which can be explained with the help of Fig.~1. Due to amplitude
dissipation, the population of lower-excited states increases, as
seen by comparing Figs.~2(a) and~4 for the two-PB, as well as
Figs.~2(b) and~5 for the three-PB.

%------------------------------------------------------------------
\begin{figure}
\includegraphics[width=.5\textwidth]{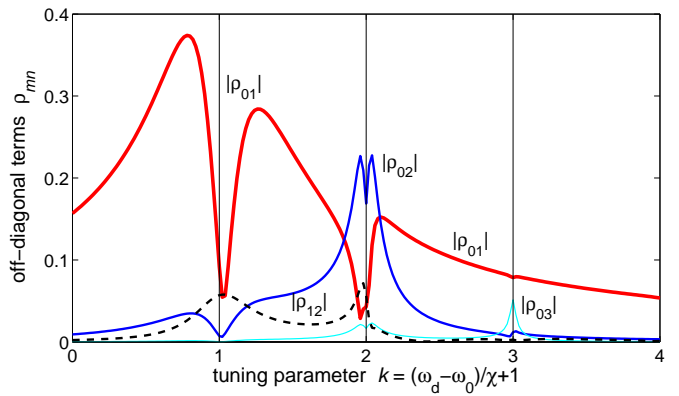}
%\vspace*{-4mm}
\caption{(Color online). Off-diagonal elements
$\rho_{nm}=\langle n|\hat\rho_{\rm ss}|m\rangle$ of the
steady-states $\hat\rho_{\rm ss}$ as a function of the tuning
parameter $k$ for the same parameters as in Fig.~2(a). These
nonzero coherences demonstrate that the steady states are not
maximally mixed even when $k$ is not an integer. }
\end{figure}

%------------------------------------------------------------------
\section{Quantum signatures of photon blockades}

%------------------------------------------------------------------
\subsection{Photon-number signatures of photon blockades}

We have already discussed some photon-number signatures of
the PB effects by analyzing Fig.~2. Now, we  focus on
demonstrating the nonclassicality of the generated steady
states. We recall that the nonclassicality (or quantumness)
of a bosonic system is usually understood if the system is
described by a nonpositive Glauber-Sudarshan quasiprobability
function~\cite{VogelBook} or, equivalently, by a negative
normally-ordered matrix of moments~\cite{Vogel08,Miran10}. In
particular, the nonclassicality can often be revealed by
analyzing photon-number properties only.

Thus, in Fig.~2, we plotted the Fano factor, which is defined
by~\cite{VogelBook}:
\begin{equation}
F(\hat{\rho}_{{\rm ss}})=\frac{\langle \hat n^{2}\rangle
-\langle\hat n\rangle ^{2}}{\langle\hat n\rangle}.\label{Fano}
\end{equation}
A given state exhibits sub-Poisson (super-Poisson) photon-number
statistics if $F<1$ ($F>1$). The sub-Poisson statistics is a
nonclassical effect.

It is seen that the generated state exhibits the sub-Poisson
photon-number statistics for $k<1.5$ and near $k=2$ as described
by the Fano factors for the parameters chosen in Fig.~2(a). Global
and local minima of the sub-Poisson statistics occur for the
single-PB ($k=1$) and two-PB ($k=2$), respectively. Note the
occurrence of local maxima of the Fano factors at points
relatively close to $k=2$. There are also local maxima at $k=3$
and 4 for the parameters $\epsilon,\,\chi$, and $\gamma$ chosen in
Fig.~2(a). However, for the approximately twice-larger driving
strength $\epsilon$ as in, e.g., Fig.~2(b), we can also observe at
$k=3$ a local minimum below 1 of the Fano factor, which
corresponds to the sub-Poisson statistics of the three-PB.

%------------------------------------------------------------------
\begin{figure}

\includegraphics[width=.45\textwidth]{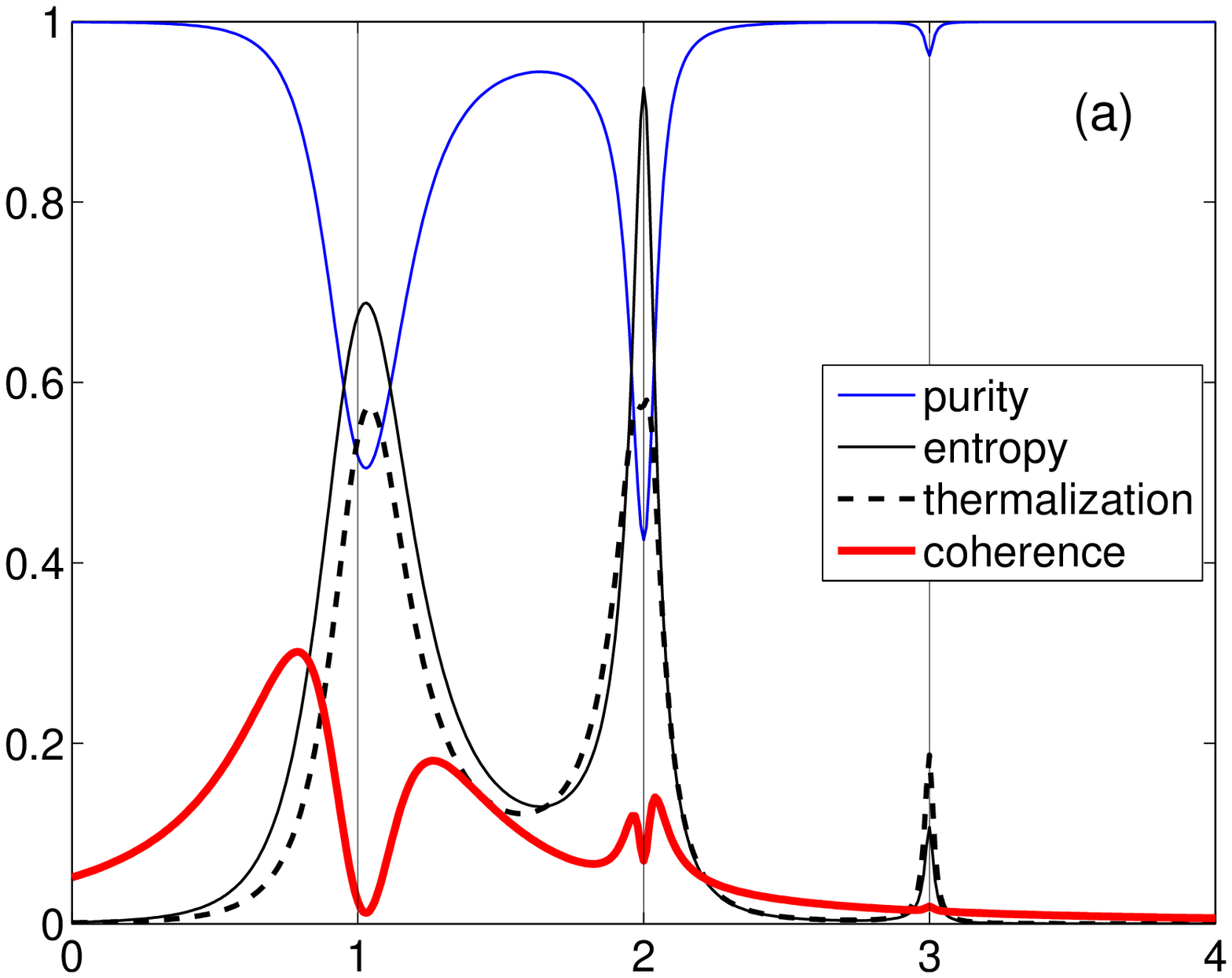}

%\vspace*{-6mm}

\includegraphics[width=.45\textwidth]{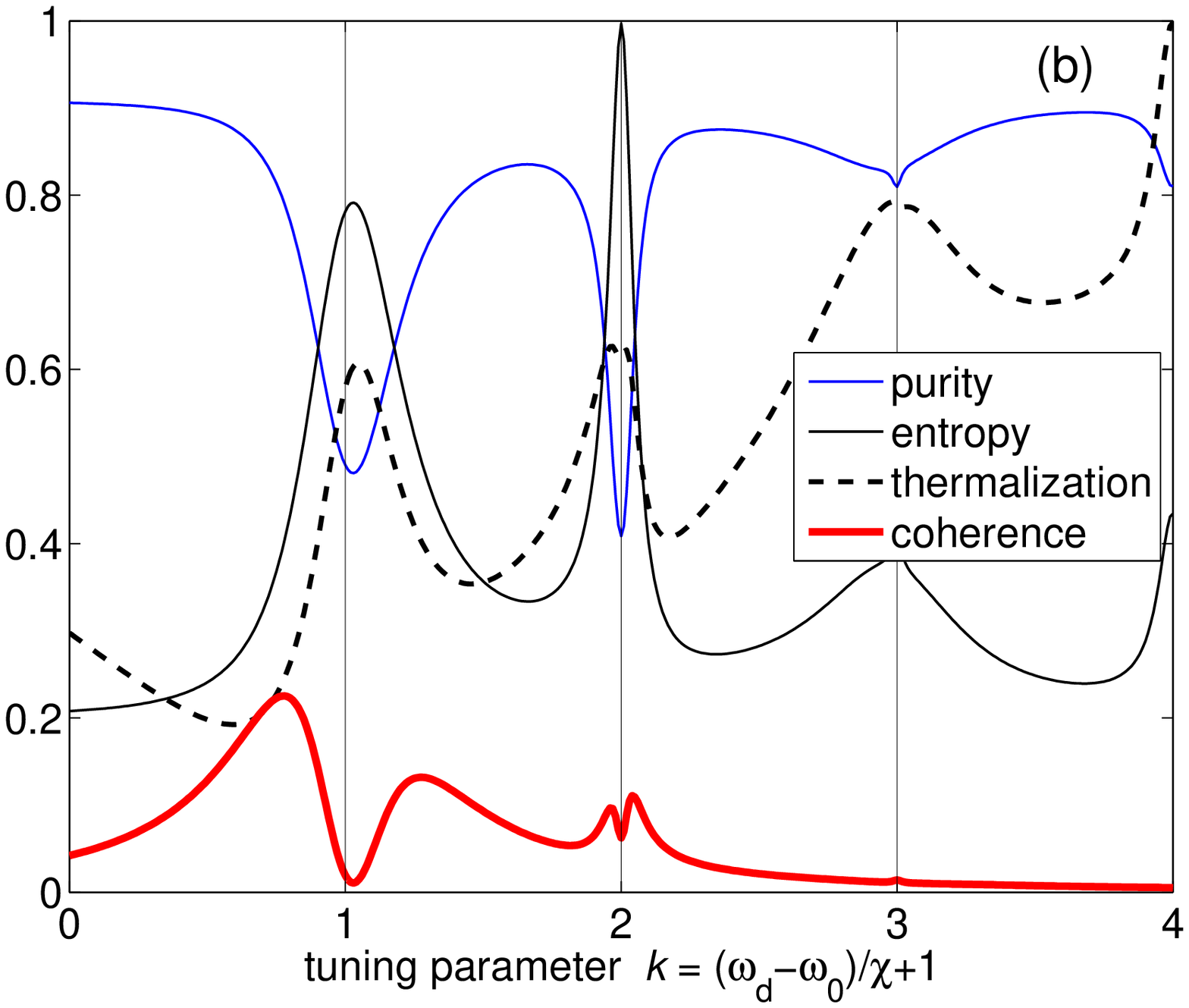}

\vspace*{-4mm} \caption{(Color online). The purity
$\mu(\hat{\rho}_{\rm ss})$, von Neumann entropy
$S(\hat\rho_{\rm ss})$, thermalization $T_{\rm
therm}(\hat\rho_{\rm ss})$, and coherence ${C}(\hat\rho_{\rm
ss})$ as a function of the tuning parameter $k$ for the mean
thermal-photon numbers (a) $\bar{n}_{\rm th}=0$ and (b)
$\bar{n}_{\rm th}=0.05$ with other parameters chosen the same
as in Fig.~2(a). This figure shows how thermal photons
strongly affect $\mu$, $S$, and $T_{\rm therm}$, but do not
affect much $C$.}
\end{figure}
%------------------------------------------------------------------
\subsection{Coherence and entropic signatures of photon blockades}

In order to show how the steady-state solutions $\hat{\rho}_{{\rm
ss}}$ depend on the tuning parameter $k$, we also analyze their
coherence properties.

Off-diagonal elements $\rho_{nm}=\langle n|\hat\rho_{\rm
ss}|m\rangle$ (for $n\neq m)$ of the density matrix $\hat\rho_{\rm
ss}$, which are also often called coherences, are shown in Fig.~6.
It is seen that almost all plotted coherences do not vanish in the
steady states even for off-resonances (i.e., when the tuning
parameter $k$ is not an integer). This means that the generated
steady states are not completely mixed.

Our deeper analysis of the coherence properties of the steady
states presented in Fig.~7 includes the following measures: the
purity $\mu(\hat{\rho}_{{\rm ss}})={\tr}(\hat{\rho}_{{\rm
ss}}^{2})$ and the coherence parameter  $C(\hat{\rho}_{{\rm ss}})$
defined as the sum of all off-diagonal terms of the density matrix
$\hat{\rho}_{{\rm ss}}$ (see, e.g.,~\cite{Dodonov00}):
\begin{eqnarray}
C(\hat{\rho}_{{\rm ss}})&=&\sum_{n\neq
m}|\rho_{nm}|^{2}={\tr}[(\hat{\rho}_{{\rm ss}}-\hat{\rho}_{{\rm
diag}})^{2}]\nonumber \\ &=&\mu(\hat{\rho}_{{\rm
ss}})-\mu(\hat{\rho}_{{\rm diag}}),\label{N1}
\end{eqnarray}
where $\hat{\rho}_{{\rm diag}}=\sum_{n}\rho_{nn} |n\rangle\langle
n|$. Thus, this parameter is just the total-state purity after
subtracting the diagonal-state purity.  It is worth noting that
$C$ is sometimes additionally normalized~\cite{Dodonov00}. It is
seen that decoherence, understood as the vanishing of the
off-diagonal elements of a density matrix, can intuitively be
quantified by $C(\hat{\rho}_{\rm ss})$.

To describe the mixedness of the generated steady state, we
calculate the von Neumann entropy $S(\hat{\rho}_{{\rm
ss}})=-\tr(\hat{\rho}_{{\rm ss}}\ln\hat{\rho}_{{\rm ss}})$ and the
thermalization parameter defined for a finite-dimensional system
as~\cite{Dodonov00}:
\begin{eqnarray}
 T(\hat{\rho}_{\rm ss})&=&\frac{S_{L}(\hat{\rho}_{\rm
 ss})}{\sqrt{\tr[(\hat{\rho}-\hat{\rho}_{0})^2]
 \tr[(\hat{\rho}-\hat{\rho}_{\max})^2]}}
  \\ &=&\frac{S_{L}(\hat{\rho}_{\rm ss})}{\sqrt{[1+\mu(\hat{\rho}_{{\rm
ss}})-2p_{0}][1+\mu(\hat{\rho}_{{\rm ss}})-2p_{\max}]}},
  \nonumber
\label{thermolization}
\end{eqnarray}
which is defined as a properly normalized linear entropy
$S_{L}(\hat{\rho}_{{\rm ss}})$. Here,
$\hat{\rho}_{0}=|0\rangle\langle0|$,
$\hat{\rho}_{\max}=|n_{\max}\rangle\langle n_{\max}|$,
$p_{0}=\langle0|\hat{\rho}_{{\rm ss}}|0\rangle$,
$p_{\max}=\langle n_{\max}|\hat{\rho}_{{\rm
ss}}|n_{\max}\rangle$, and $|n_{\max}\rangle$ is the
uppermost Fock state generated in the system. As explained in
Ref.~\cite{Dodonov00}, this normalization of the linear
entropy is done to exclude somehow the contribution of the
ground state $\hat{\rho}_{0}=|0\rangle\langle0|$, which has a
double nature: It is both a pure state and a completely
decoherent equilibrium state (as in our model without the
driving force). Moreover, the linear entropy
$S_{L}(\hat{\rho}_{{\rm ss}})$, which is another parameter of
mixedness, can easily be obtained from the purity (depicted
in Fig.~7), since $S_{L}(\hat{\rho}_{{\rm
ss}})=1-\mu(\hat{\rho}_{{\rm ss}})$. Other aspects of the
quantum entropies and mixedness in the discussed model (but
only in the special case of $k=1$) were studied in
Ref.~\cite{Bajer04}.

One can clearly see in Fig.~7 that the entropy
$S(\hat{\rho}_{{\rm ss}})$ and thermalization $T_{\rm
therm}(\hat{\rho}_{{\rm ss}})$ reach maxima and,
equivalently, the purity $\mu(\hat{\rho}_{{\rm ss}})$ and
coherence $C(\hat{\rho}_{{\rm ss}})$ parameters have minima
at (or very close to) $k=1$, 2, 3. It is worth comparing this
behavior with the Fano factor $F(\hat{\rho}_{{\rm ss}})$
shown in Fig.~2(a) for the same parameters, where maxima are
observed only for $k=1$ and 2 (where the PBs are predicted),
while for $k=3$ the minimum occurs, which corresponds to the
case where the PB is practically not observed for the chosen
parameters, as we conclude by analyzing the probabilities
$P_{n}$ shown in Fig.~2(a). By contrast, the coherence
parameter $C(\hat{\rho})$ has minima at $k\approx 1,2$ and
very small maximum at $k=3$. By comparing Figs.~7(a) and 7(b)
obtained for the system coupled to a zero- and
nonzero-temperature reservoirs, we can easily see how the PBs
are sensitive to temperature or, equivalently, to the number
$\bar{n}_{\rm th}$ of thermal photons. Even by adding a very
small number of thermal photons, such as $\bar{n}_{\rm
th}=0.05$, the linear and von Neumann entropies together with
the thermalization parameter are noticeably increased. This
is not the case for the coherence parameter
$C(\hat{\rho}_{{\rm ss}})$.

%------------------------------------------------------------------
\begin{figure}

\includegraphics[width=.5\textwidth]{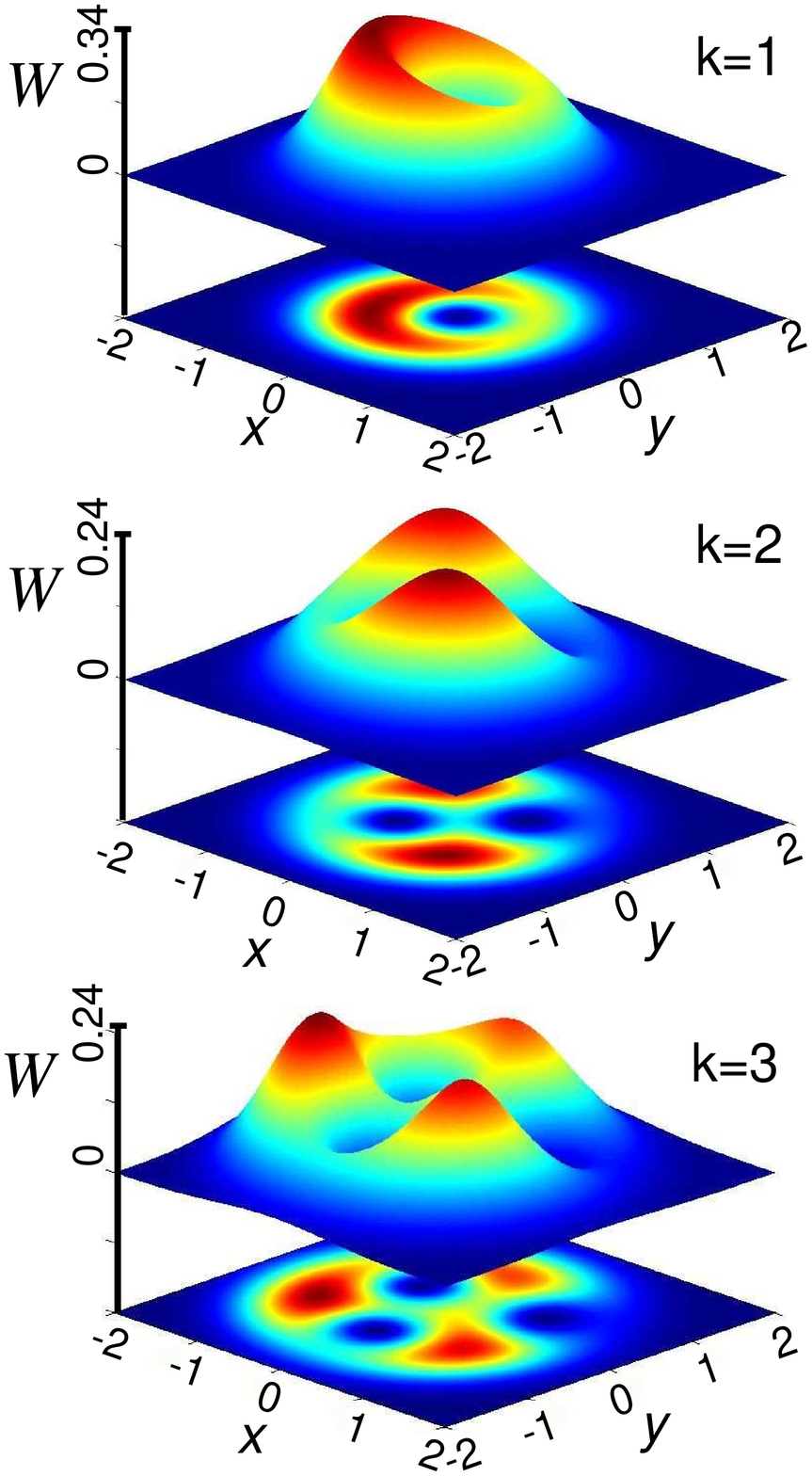}

\vspace*{-8mm} \caption{(Color online). Wigner functions
$W(\alpha=x+ip)$ for the $k$-photon blockades corresponding
to the steady-state solutions of the master equation for the
Hamiltonian $\hat{H}_{\mathrm{rot}}^{(k)}(0)$, with the same
parameters as in Fig.~2(a) except the driving strength
$\epsilon=11.56\gamma$ for $k=3$. It is seen that the Wigner
function for the $k$-photon blockade has $k$ peaks and $k$
dips. For $k=3$ the peaks are deformed (due to interference
in phase space) but still are visible. }
\end{figure}
%------------------------------------------------------------------
\subsection{Phase-space description of photon-blockades}

The dynamics of quantum systems can equivalently be described by
using the phase-space formalism of Wigner functions. This
formalism is particularly useful for distinguishing different PBs
effects as we show below. The Wigner function can be defined
as~\cite{Cahill69}:
\begin{equation}
W(\alpha)=\frac 2\pi{\tr}\big[\hat D^{-1}(\alpha)\hat \rho\hat
D(\alpha) \hat P \big], \label{Wigner}
\end{equation}
where $\hat D(\alpha)=\exp(\alpha\hat a^\dagger-\alpha^*\hat a)$
is the displacement operator with a complex number $\alpha$, and
$\hat P=\exp(i\pi\hat a^\dagger \hat a)$ is the parity operator,
so its  action on Fock states is simply given by $\hat
P|n\rangle=(-1)^n|n\rangle$. The Wigner function can be
generalized to the $s$-parametrized Cahill-Glauber
quasiprobabilities~\cite{Cahill69}. Nevertheless, contrary to
other definitions, Eq.~(\ref{Wigner}) shows a \emph{direct} method
(i.e., without the necessity of applying quantum state tomography)
to measure the Wigner function~\cite{Lutterbach97}. This direct
method was experimentally applied in, e.g., cavity
QED~\cite{Bertet02} and circuit QED~\cite{Hofheinz09} systems.

In Fig.~8, we show the Wigner function for the blockades of up to
$k$ photons for $k=1,2,3$ by properly choosing parameters in order
to satisfy the resonance condition $\Delta_{k}=0$ together with
$\gamma\ll\epsilon\ll\chi$. The $k$-peak and $k$-dip (anti-peak)
structures of the Wigner functions clearly correspond to the
$k$-PB.

%------------------------------------------------------------------
\begin{figure}

\includegraphics[width=.222\textwidth]{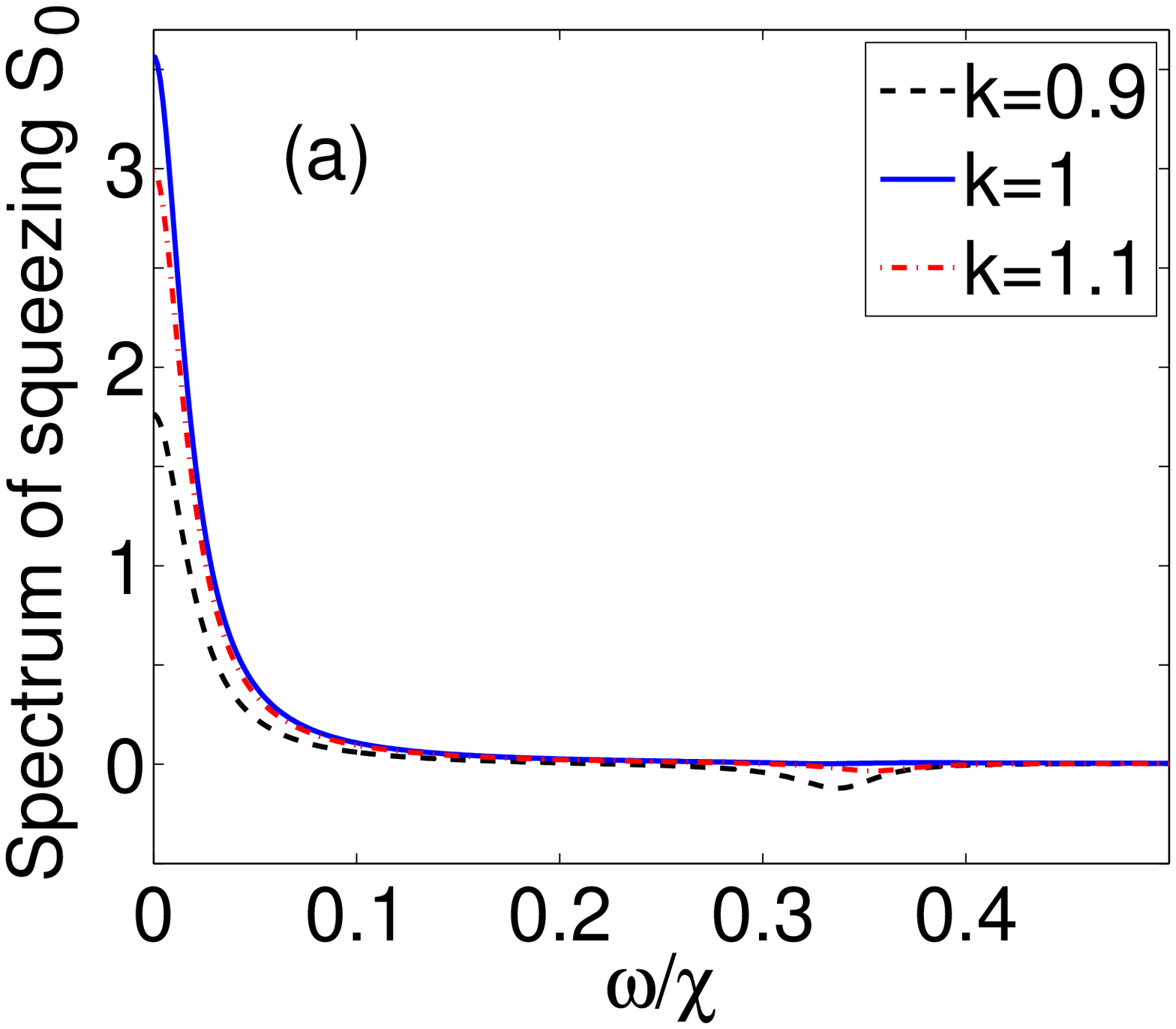}
% \vspace*{-1mm}
\includegraphics[width=.25\textwidth]{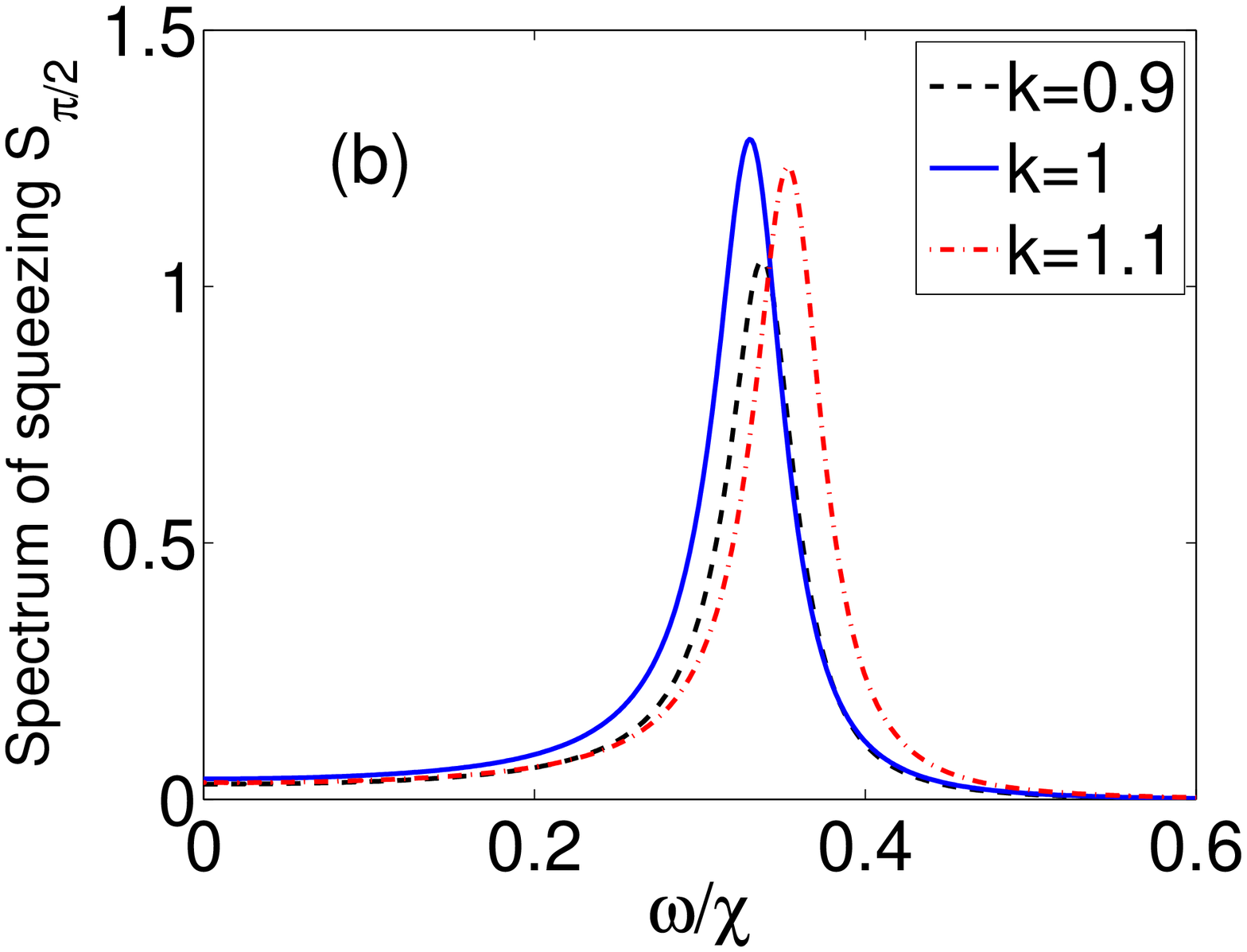}

\vspace*{-4mm} \caption{(Color online). Spectra of squeezing
(a) $S_{0}(\omega)$ and (b) $S_{\pi/2}(\omega)$ for the
single-photon blockade for the same parameters as in
Fig.~2(a). Broken curves show the spectra for the Hamiltonian
$\hat{H}_{\mathrm{rot}}^{(k)}(0)$ with off-resonance values
of the tuning parameter $k$.}
\end{figure}

%------------------------------------------------------------------
\begin{figure}

\hspace*{-4mm}\includegraphics[width=.25\textwidth]{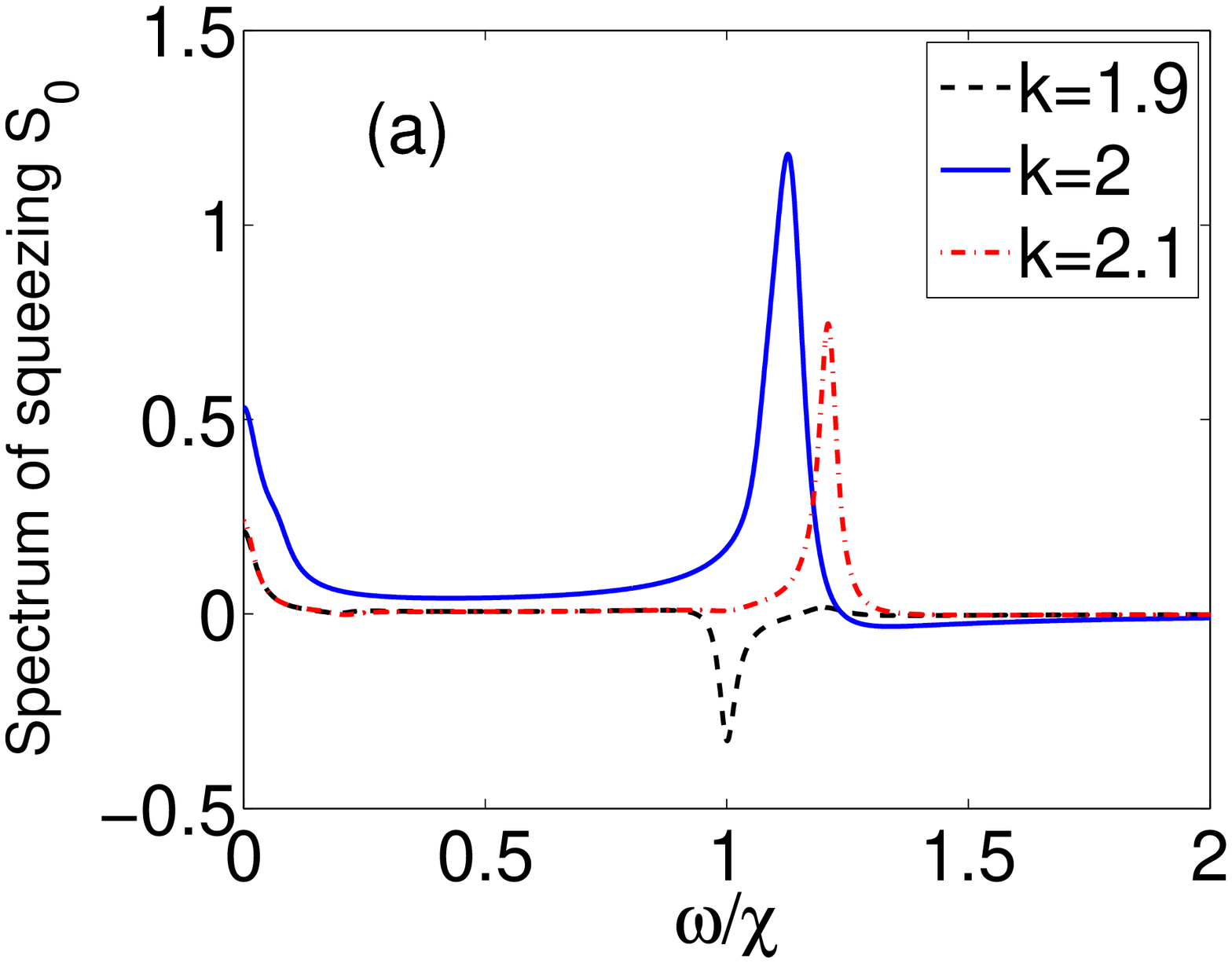}
%\vspace*{-1mm}
\includegraphics[width=.25\textwidth]{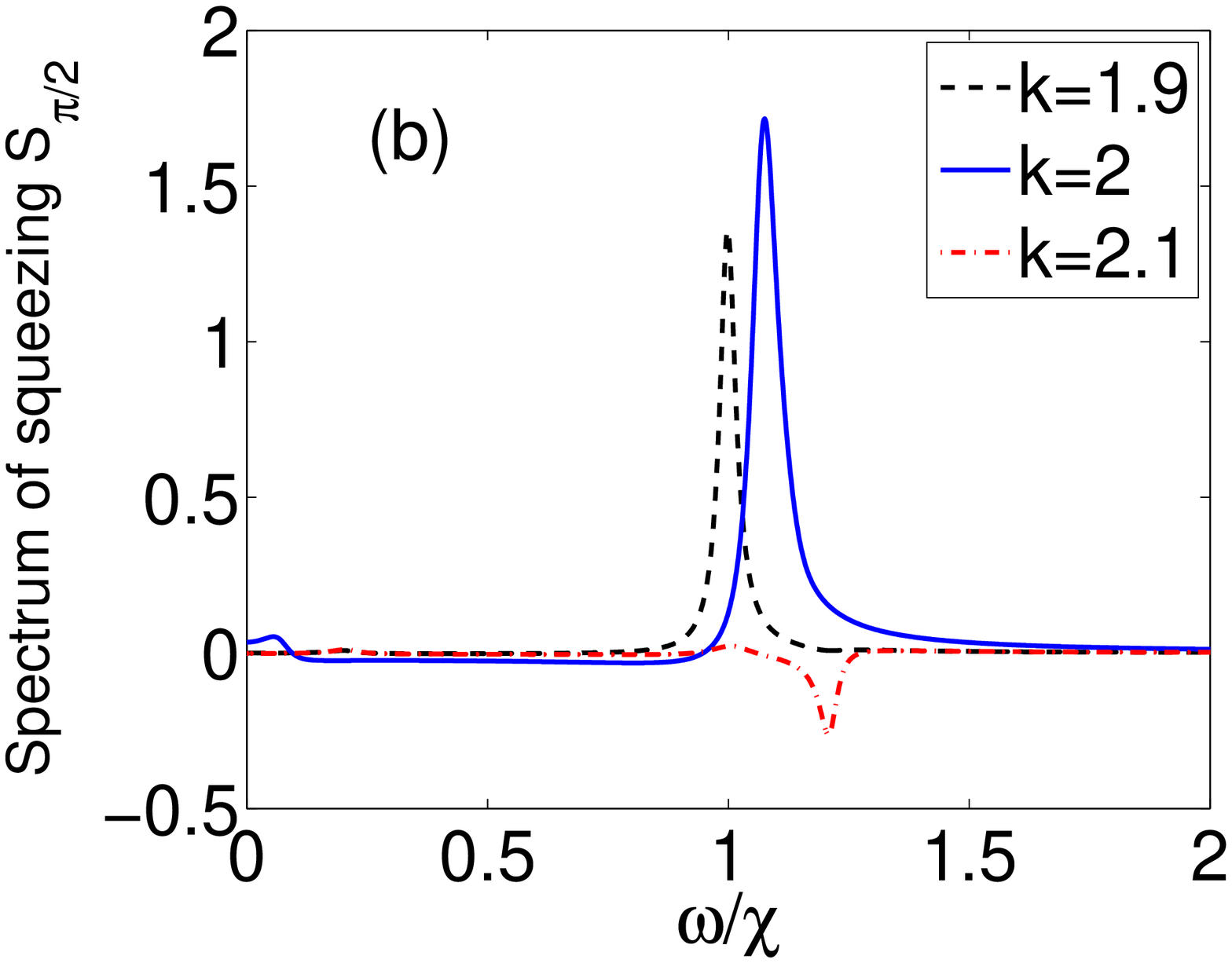}

\vspace*{-4mm} \caption{(Color online). Spectra of squeezing
for the two-photon blockade analogously to those in Fig.~9
but for the resonance $k=2$ and off-resonance values of the
tuning parameter $k$. It is seen that either $S_{0}$ or
$S_{\pi/2}$ only for the off-resonant cases has a clear
negative dip.}
\end{figure}

%------------------------------------------------------------------
\begin{figure}

\includegraphics[width=.237\textwidth]{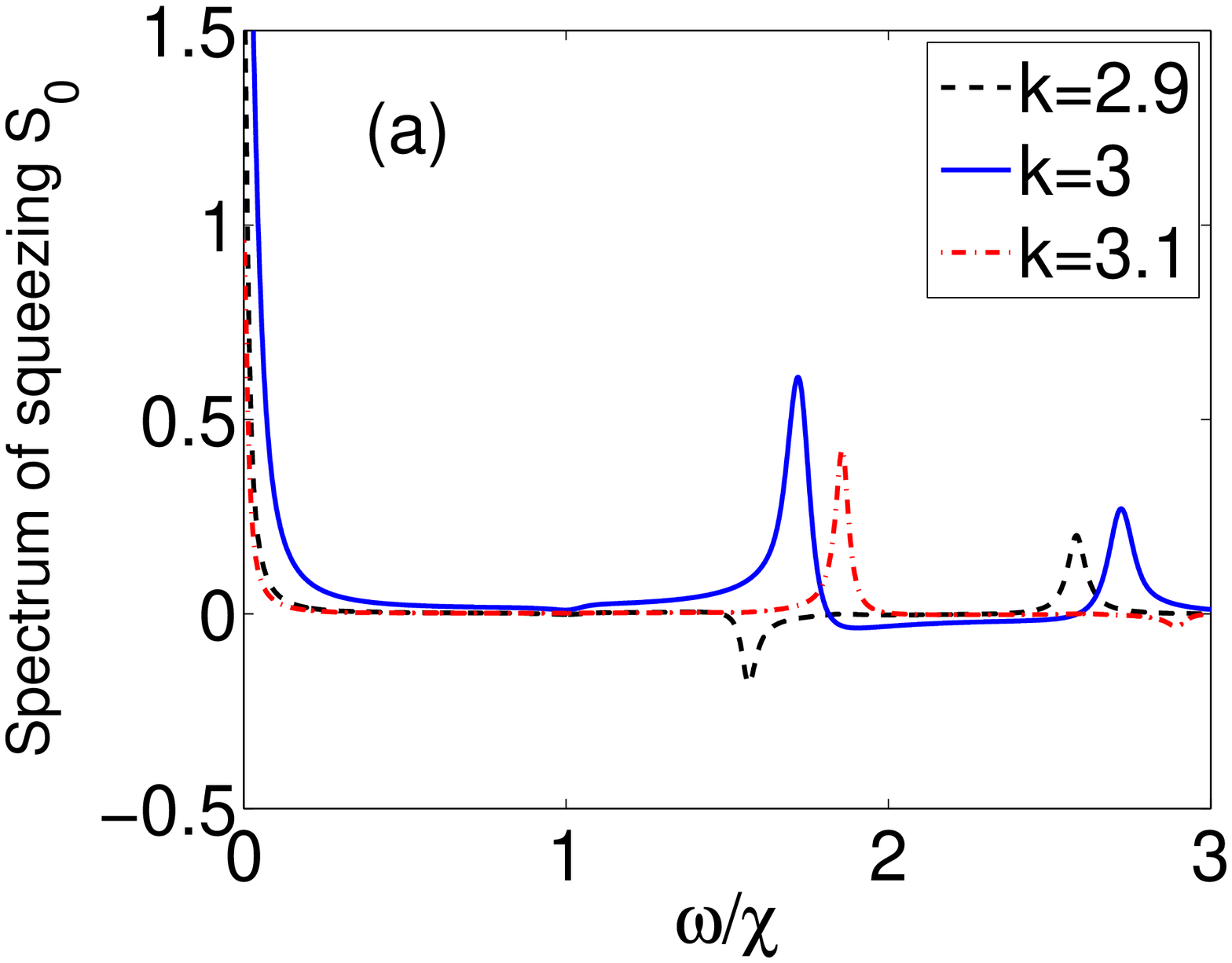}
% \vspace*{-1mm}
\includegraphics[width=.237\textwidth]{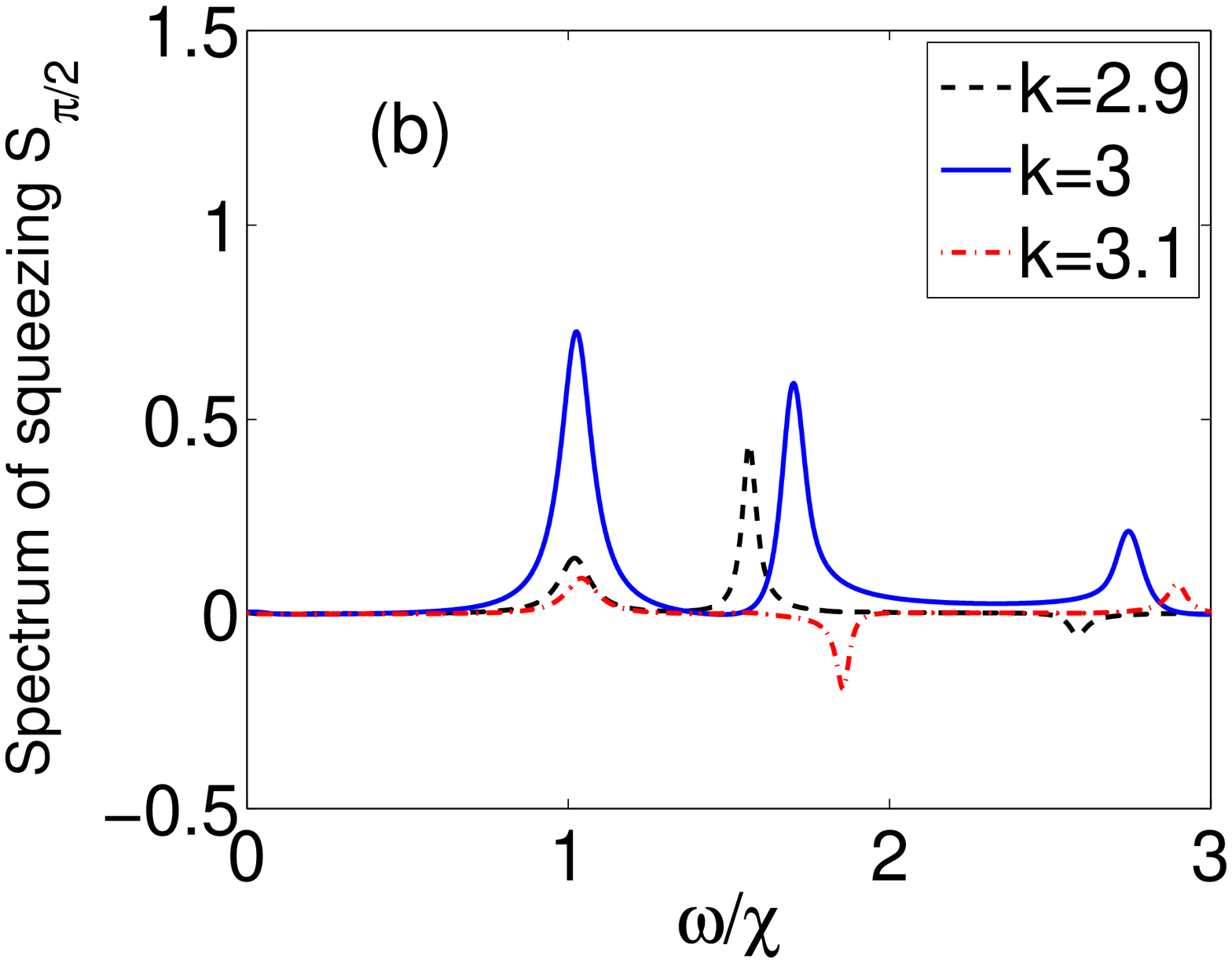}

\vspace*{-4mm} \caption{(Color online). Spectra of squeezing
for the three-photon blockade analogous to those in
Figs.~9-10 but for the resonance $k=3$ and off-resonance
values of the tuning parameter $k$ (assuming the driving
strength $\epsilon=11.56\gamma$). Analogously to the
two-photon blockade, the off-resonance spectra have a
negative dip either for $\theta=0$ or $\pi/2$, which does not
appear for the resonant case for $k=3$. }
\end{figure}
%------------------------------------------------------------------
\subsection{Spectrum of squeezing for photon-blockades}

The PBs can also be revealed in two-time correlations. Let us
analyze the two-time normally-ordered ($::$) and time-ordered
$({\cal T})$ correlation function
\begin{eqnarray}
{\cal T}\langle :\hat X_\theta(\tau)\hat X_\theta(0):\rangle\equiv
\lim_{t\rightarrow\infty}{\cal T}\langle :
\hat X_\theta(t+\tau)\hat X_\theta(t):\rangle
\nonumber \\
= \langle\hat a^\dagger(\tau)a(0) \rangle +\langle\hat
a^\dagger(0)a(\tau) \rangle \hspace{13mm}
\nonumber \\
+\;e^{-2i\theta} \langle\hat a(\tau)a(0) \rangle +e^{2i\theta}
\langle\hat a^\dagger(0)a^\dagger(\tau) \rangle
\end{eqnarray}
of the quadrature phase distribution
\begin{equation}
\hat X_\theta(t)=\hat a(t)e^{-i\theta}+\hat
a^\dagger(t)e^{i\theta}. \label{N12}
\end{equation}
The spectrum of squeezing is defined as the Fourier transform of
the covariance~\cite{WallsBook}:
\begin{equation}
S_{\theta}(\omega)=\int_{-\infty}^{\infty}d \tau
e^{-i\omega\tau}{\cal T}\langle
:\hat X_\theta(\tau),\hat X_\theta(0):\rangle, \label{N13}
\end{equation}
where the covariance is defined by the general formula
$\left\langle A,B\right\rangle =\left\langle AB\right\rangle
-\left\langle A\right\rangle \left\langle B\right\rangle$. In
Figs.~9-11, we demonstrated distinctive properties of the spectra
of squeezing for the single-PB (Fig.~9), two-PB (Fig.~10) and
three-PB (Fig.~11) for the tuning parameters at the resonance,
$k=k_0$, and slightly out of the resonance, $k=k_0\pm\delta$,
where $\delta\ll 1$ and $k_0=1,2,3$. Surprisingly, the spectra of
squeezing as a function of the tuning parameter $k$ are more
sensitive indicators of the two- and three-PBs instead of those of
the single-PB.

%------------------------------------------------------------------
\section{Conclusions}

We studied nonlinear photon-photon interaction at the
single-photon level in a system consisting of a cavity with a Kerr
nonlinearity driven by a weak classical  field.  This is a
standard prototype model, where the Kerr-like nonlinearity can be
induced by the interaction of, e.g., a qubit with a cavity field
in the dispersive regime.

By finding the master-equation solutions of the model in the
steady-state limit for a properly chosen frequency of the
classical driving field, we observed a blockade of more than one
photon. This is a generalization of the single-PB for the
multiphoton (say $m$-photon) case, which means that the Fock
states with $n=0,1,...,m$ photons are, practically, the
\emph{only} generated in this nonlinear cavity. This effect can
also be interpreted as the multiphoton-state truncation realized
by nonlinear quantum scissors or multiphoton-induced tunneling
corresponding to multi-photon transitions as schematically shown
in Fig.~1.

It is worth noting that some generalizations of the
single-photon nonlinear truncation processes for the
multiphoton cases were already discussed by, e.g., Leo\'nski
\etal~(see Ref.~\cite{Miran01} for a review). Nevertheless,
all these proposals assume either higher-order driving
processes, as described in Ref.~\cite{Leonski96} by the
Hamiltonian $\hat H^{(l)}_{\rm drive}
=\epsilon^{(l)}[\hat{a}^l+(\hat{a}^{\dagger})^l],$ or
higher-order Kerr nonlinearity,  corresponding to the
Hamiltonian $\hat H^{(k)}_{\rm Kerr}
=\chi^{(k+1)}(\hat{a}^{\dagger})^{k+1} \hat{a}^{k+1}$ in
Ref.~\cite{Leonski97}, or both $\epsilon^{(k)}$ and
$\chi^{(l)}$~\cite{Miran96} for $k,l=2,3,...$. In our
approach, we assume the lowest-order parametric driving
($l=1$) and the lowest-order Kerr nonlinearity ($k=1$), which
is the same as in the standard single-PB. Also, in the
mentioned generalizations, only the free evolution was
analyzed. Thus, these effects can solely be interpreted as
nonstationary-field effects; so, they are almost
unmeasurable. The dissipation effects were later studied, but
only for the short-time evolution regime in
Ref.~\cite{Miran04}. In this paper, we  focused on measurable
effects, thus, we studied (except only one section) the
steady-state solutions of the master equation in the
infinite-time limit.

We described, as given by Eq.~(\ref{k}),  how to choose the
resonance frequency $\omega_{d}$ of the driving field for a given
cavity frequency $\omega_0$ and the Kerr nonlinearity $\chi$ and
how to increase the driving strength $\epsilon$ (see Fig.~3) in
order to observe the blockade at the multiphoton Fock state.

In our presentation we focused on the blockades of  $k=2$ and 3
photons in order to achieve high fidelity $F_k$ of the $k$-photon
truncation and with a contribution of the $k$-photon state which
cannot be neglected, which means that $F_{k-1}\ll 1$.

We showed analytically and numerically that these PBs occur
both for non-dissipative and dissipative systems. In
particular, we showed an excellent agreement (i.e., no
differences on the scale of Fig.~4) for the two-PB between
the numerical solutions obtained in the Hilbert spaces of
infinite-dimension (practically $N_{\rm dim}=100$) and of
finite dimension ($N_{\rm dim}=4$). Our approximate
analytical solutions only slightly differ from the numerical
ones (as shown in Fig.~4 for a relatively large~$\delta$).

We demonstrated a variety of quantum properties revealing the
unique nature of the single-, two-, and three-PBs as a
function of the tuning parameter $k$. In particular, we
studied photon-number statistics (as shown in Fig.~2), as
well as coherence and entropic properties (in Figs.~6 and 7).
We gave a clear comparison of the Wigner functions of the
steady states corresponding to the blockades of $k=1,2,3$
photons (see Fig.~8). Moreover, we showed clear differences
in the spectra of squeezing of the steady states in the
resonant cases at $k=1,2,3$ and slightly off the resonances
(as shown in Figs.~9-11).

We suggested that the two-PB and three-PB can be observed in
various systems, where the single-PB has already been
experimentally observed~\cite{Birnbaum05,Faraon08,Lang11} or, at
least, theoretically predicted~\cite{Rabl11}.

Analogously, a
blockade of \emph{phonons} instead of \emph{photons} can be
considered. Namely,  a two-phonon generalization of the standard
single-phonon blockade can be predicted in the nanomechanical
systems studied in Refs.~\cite{Liu10,Didier11}. The crucial point
is that the effective Hamiltonians, given by Eqs.~(\ref{H0})
and~(\ref{Hk}), under proper conditions, can be obtained from
other standard models in cavity QED, circuit QED, and quantum
optomechanics.

The single-photon blockade has attracted considerable
interest, with potential applications in quantum state
engineering, quantum information, and quantum communication.
We believe that the multi-photon blockades described in this
paper can also find some useful quantum applications and in
addition, at a fundamental level, they can show a deeper
analogy between condensed-matter and optical phenomena.

%------------------------------------------------------------------
\begin{acknowledgments}
We thank Karol Bartkiewicz for discussions. A.M. and M.P.
acknowledge the support of the Polish National Science Centre
under Grants No. DEC-2011/03/B/ST2/01903 and No.
DEC-2012/04/M/ST2/00789. Y.X.L. was supported by the National
Natural Science Foundation of China Grants under No. 10975080
and No. 61025022. J.B. was supported by the Czech Ministry of
Education under Project No. MSM6198959213. F.N. acknowledges
partial support from the Army Research Office, JSPS-RFBR
Contract No.~12-02-92100, Grant-in-Aid for Scientific
Research (S), MEXT Kakenhi on Quantum Cybernetics, and
Funding Program for Innovative R\&D on S\&T (FIRST).
\end{acknowledgments}

%------------------------------------------------------------------

\end{document}